\documentclass[sigconf]{acmart}

\usepackage{epsfig,endnotes}
\usepackage[utf8]{inputenc}

\usepackage{amssymb}
\usepackage{array}
\usepackage{arydshln}
\usepackage{booktabs}
\usepackage{collcell}
\usepackage{comment}
\usepackage{filecontents}
\usepackage{float}
\usepackage{ifthenx}
\usepackage{lipsum}
\usepackage{makecell}
\usepackage{multirow}
\usepackage{nicefrac}
\usepackage{paralist}
\usepackage{upgreek}
\usepackage{siunitx}
\usepackage{todonotes}
\usepackage{threeparttable}
\PassOptionsToPackage{hyphens}{url}\usepackage{url}
\usepackage{xspace}
\usepackage{placeins}

\def\subheading#1{\medskip\noindent{\boldmath\textbf{#1}}~\ignorespaces}

\usepackage{tikz}
\usetikzlibrary{external}
\usepackage{pgfplots}
\usetikzlibrary{pgfplots.groupplots}
\usetikzlibrary{arrows}
\usetikzlibrary{patterns}
\usetikzlibrary{positioning}
\usetikzlibrary{decorations.pathreplacing}
\usetikzlibrary{shapes.arrows}
\usetikzlibrary{pgfplots.groupplots}
\pgfplotsset{compat=1.8}

\usepackage{subcaption}
\usepackage{letltxmacro}
\reversemarginpar
\usepackage{todonotes}
\usepackage{marginnote} 
\let\marginpar\marginnote
\LetLtxMacro{\oldtodo}{\todo}
\renewcommand{\todo}[2][]{\tikzexternaldisable\oldtodo[fancyline,size=\footnotesize,#1]{#2}\tikzexternalenable}
\renewcommand{\todo}[1]{\tikzexternaldisable\oldtodo[fancyline,size=\footnotesize]{#1}\tikzexternalenable}

\usepackage{algorithmic}
\usepackage[ruled]{algorithm2e}
\usepackage{listings}
\newfloat{listing}{tbhp}{lst}
\floatname{listing}{Listing}
\lstdefinelanguage
   [x64]{Assembler}     
   [x86masm]{Assembler} 
   {morekeywords={CDQE,CQO,CMPSQ,CMPXCHG16B,JRCXZ,LODSQ,MOVSXD, %
                  POPFQ,PUSHFQ,SCASQ,STOSQ,IRETQ,RDTSCP,SWAPGS, %
                  rax,rdx,rcx,rbx,rsi,rdi,rsp,rbp, %
                  r8,r8d,r8w,r8b,r9,r9d,r9w,r9b}} 

\newcommand{\gbr}[1]{\left\{#1\right\}}

\newcommand{\SIx}[1]{\num{#1}\relax}

\newcommand{\etal}{et~al.\ } 
\newcommand{\ie}{\textit{i.e.},\ } 
\newcommand{\eg}{e.g.,\ } 
\newcommand{\cf}{cf.\ } 
\newcommand{\KeyDrown}{\emph{KeyDrown}\xspace}

\newcommand{\FlushReload}{Flush+\allowbreak Reload\xspace}

\newcommand{\PrimeProbe}{Prime+\allowbreak Probe\xspace}
\newcommand{\MultiPrimeProbe}{Multi-\allowbreak Prime+\allowbreak Probe\xspace}

\newcommand{\GTKplus}{\emph{GTK+}\xspace}

\newcommand{\OnePlus}{OnePlus 3T\xspace}
\newcommand{\Nexus}{Nexus 5\xspace}
\newcommand{\FScore}{F-score\xspace}

\newcommand{\ReqOne}{\emph{R1}\xspace}
\newcommand{\ReqTwo}{\emph{R2}\xspace}
\newcommand{\ReqThree}{\emph{R3}\xspace}

\usepackage{pifont}
\newcommand{\cmark}{\leavevmode{\color{green}\ding{51}}}%
\mathchardef\mhyphen="2D
\newcommand*\circled[1]{\tikz[baseline=(char.base)]{
            \node[shape=circle,draw,inner sep=2pt] (char) {#1};}}
\newcommand*\circleds[1]{\tikz[baseline=(char.base)]{
            \node[shape=circle,draw,inner sep=1pt] (char) {\footnotesize #1};}}

\newcommand{\FakeMarker}{{\color{red}$\blacktriangle$}}
\newcommand{\RealMarker}{{\color{green}$\bullet$}}

\usepackage{epsdice}
\usepackage{enumitem}

\definecolor{TolDarkPurple}{HTML}{332288}
\definecolor{TolDarkBlue}{HTML}{6699CC}
\definecolor{TolLightBlue}{HTML}{88CCEE}
\definecolor{TolLightGreen}{HTML}{44AA99}
\definecolor{TolDarkGreen}{HTML}{117733}
\definecolor{TolDarkBrown}{HTML}{999933}
\definecolor{TolLightBrown}{HTML}{DDCC77}
\definecolor{TolDarkRed}{HTML}{661100}
\definecolor{TolLightRed}{HTML}{CC6677}
\definecolor{TolLightPink}{HTML}{AA4466}
\definecolor{TolDarkPink}{HTML}{882255}
\definecolor{TolLightPurple}{HTML}{AA4499}

\definecolor{PlotColorBlue}{HTML}{2C7FB8}
\definecolor{PlotColorRed}{HTML}{F03B20}
\definecolor{PlotColorGreen}{HTML}{31A354}

\definecolor{red}{HTML}{F03B20}
\definecolor{yellow}{HTML}{F5EE9A}
\definecolor{green}{HTML}{BEDB39}
\definecolor{blue}{HTML}{2C7FB8}

\pgfplotscreateplotcyclelist{mbarplot cycle}{%
  {draw=TolDarkBlue,    fill=TolDarkBlue!70},
  {draw=TolLightBrown,  fill=TolLightBrown!70},
  {draw=TolLightGreen,  fill=TolLightGreen!70},
  {draw=TolDarkPink,    fill=TolDarkPink!70},
  {draw=TolDarkPurple,  fill=TolDarkPurple!70},
  {draw=TolDarkRed,     fill=TolDarkRed!70},
  {draw=TolDarkBrown,   fill=TolDarkBrown!70},
  {draw=TolLightRed,    fill=TolLightRed!70},
  {draw=TolLightPink,   fill=TolLightPink!70},
  {draw=TolLightPurple, fill=TolLightPurple!70},
  {draw=TolLightBlue,   fill=TolLightBlue!70},
  {draw=TolDarkGreen,   fill=TolDarkGreen!70},
}

\pgfplotscreateplotcyclelist{mlineplot cycle}{%
  {TolDarkBlue, mark=*, mark size=1.5pt},
  {TolLightBrown, mark=square*, mark size=1.3pt},
  {TolLightGreen, mark=triangle*, mark size=1.5pt},
  {TolDarkBrown, mark=diamond*, mark size=1.5pt},
}

\pgfplotsset{
  compat=1.9
}
\pgfplotsset{
  mlineplot/.style={
  },
}

\pgfplotsset{
  mbarplot base/.style={
    mbaseplot,
    bar width=6pt,
    axis y line*=none,
  },
}

\pgfplotsset{
  mbarplot/.style={
    mbarplot base,
    ybar,
    xmajorgrids=false,
    ymajorgrids=true,
    area legend,
    legend image code/.code={%
      \draw[#1] (0cm,-0.1cm) rectangle (0.15cm,0.1cm);
    },
    cycle list name=mbarplot cycle,
  },
}

\pgfplotsset{
  horizontal mbarplot/.style={
    mbarplot base,
    xmajorgrids=true,
    ymajorgrids=false,
    xbar stacked,
    area legend,
    legend image code/.code={%
      \draw[#1] (0cm,-0.1cm) rectangle (0.15cm,0.1cm);
    },
    cycle list name=mbarplot cycle,
  },
}

\pgfplotsset{
  mbaseplot/.style={
    legend style={
      draw=none,
      fill=none,
      cells={anchor=west},
    },
    x tick label style={
      font=\footnotesize
    },
    y tick label style={
      font=\footnotesize
    },
    legend style={
      font=\footnotesize
    },
    major grid style={
      dotted,
    },
    axis x line*=bottom,
  },
  disable thousands separator/.style={
    /pgf/number format/.cd,
      1000 sep={}
  },
}

\newcommand{\ValMinTraces}{1825}

\newcommand{\ValReqTraces}{2458}

\newcommand{\ValFscoreOneOracle}{0.15}
\newcommand{\ValFscoreRandOracle}{0.14}

\newcommand{\ValFscoreRistenpart}{0.96}
\newcommand{\ValFscoreGruss}{0.96}

\newcommand{\ValPrecisionFR}{1.00}

\newcommand{\ValPrecisionFRKD}{0.05}
\newcommand{\ValFscoreFR}{0.99}
\newcommand{\ValFscoreFRKD}{0.09}

\newcommand{\ValPrecisionPP}{0.71}
\newcommand{\ValRecallPP}{0.92}
\newcommand{\ValPrecisionPPKD}{0.06}
\newcommand{\ValFscorePP}{0.81}
\newcommand{\ValFscorePPKD}{0.11}
\newcommand{\ValPPCacheSets}{5}

\newcommand{\ValPrecisionPPKDUser}{0.05}

\newcommand{\ValFscorePPKDUser}{0.10}

\newcommand{\ValPrecisionRDTSC}{0.89}
\newcommand{\ValPrecisionRDTSCKD}{0.07}
\newcommand{\ValFscoreRDTSC}{0.94}
\newcommand{\ValFscoreRDTSCKD}{0.14}

\newcommand{\ValPrecisionProc}{1.00}
\newcommand{\ValPrecisionProcKD}{0.08}
\newcommand{\ValFscoreProc}{1.00}
\newcommand{\ValFscoreProcKD}{0.15}

\newcommand{\ValPrecisionFROP}{0.88}

\newcommand{\ValPrecisionFRKDOP}{0.05}
\newcommand{\ValFscoreFROP}{0.93}
\newcommand{\ValFscoreFRKDOP}{0.10}

\newcommand{\ValPrecisionPPOP}{0.80}
\newcommand{\ValRecallPPOP}{1.00}
\newcommand{\ValPrecisionPPKDOP}{0.10}
\newcommand{\ValFscorePPOP}{0.89}
\newcommand{\ValFscorePPKDOP}{0.07}
\newcommand{\ValPPCacheSetsOP}{5}

\newcommand{\ValPrecisionRDTSCOP}{0.99}

\newcommand{\ValFscoreRDTSCOP}{0.99}
\newcommand{\ValFscoreRDTSCKDOP}{0.15}

\newcommand{\ValPrecisionProcOP}{1.00}

\newcommand{\ValFscoreProcOP}{1.00}
\newcommand{\ValFscoreProcKDOP}{0.15}

\newcommand{\ValPrecisionFRNexus}{1.00}

\newcommand{\ValPrecisionFRKDNexus}{0.01}
\newcommand{\ValFscoreFRNexus}{0.99}
\newcommand{\ValFscoreFRKDNexus}{0.02}

\newcommand{\ValPrecisionPPNexus}{0.71}
\newcommand{\ValRecallPPNexus}{0.92}
\newcommand{\ValPrecisionPPKDNexus}{0.06}
\newcommand{\ValFscorePPNexus}{0.80}
\newcommand{\ValFscorePPKDNexus}{0.11}
\newcommand{\ValPPCacheSetsNexus}{5}

\newcommand{\ValPrecisionRDTSCNexus}{0.89}

\newcommand{\ValFscoreRDTSCNexus}{0.94}
\newcommand{\ValFscoreRDTSCKDNexus}{0.14}

\newcommand{\ValPrecisionProcNexus}{1.00}

\newcommand{\ValFscoreProcNexus}{1.00}
\newcommand{\ValFscoreProcKDNexus}{0.15}

\newcommand{\ValLmbenchLatency}{6.9}
\newcommand{\ValPARSEC}{2.5}
\newcommand{\ValPARSECOne}{2.0}
\newcommand{\ValPARSECTwo}{2.5}
\newcommand{\ValPARSECFour}{3.1}

\newcommand{\ValBatteryIdle}{3.9}
\newcommand{\ValBatteryKeyboard}{15.6}

\begin{document}

\title{KeyDrown: Eliminating Keystroke Timing Side-Channel Attacks}

\author{
Michael Schwarz, Moritz Lipp, Daniel Gruss, Samuel Weiser, \\ Clémentine Maurice, Raphael Spreitzer, and Stefan Mangard\\
Graz University of Technology, Austria 
}

\renewcommand{\shortauthors}{M. Schwarz et al.}

\acmDOI{XX.XXX/XXX_X}

\acmISBN{XXX-XXXX-XX-XXX/XX/XX}

\acmConference[Pre-print]{}{arXiv}{2017}
\acmYear{2017}
\copyrightyear{2017}
\acmPrice{15.00}


\begin{abstract}
Besides cryptographic secrets, side-channel attacks also leak sensitive user input.
The most accurate attacks exploit cache timings or interrupt information to monitor keystroke timings and subsequently infer typed words and sentences. 
Previously proposed countermeasures fail to prevent keystroke timing attacks as they 
do not protect keystroke processing among the entire software stack. 

We close this gap with \KeyDrown, a new defense mechanism against keystroke timing attacks.
\KeyDrown injects a large number of fake keystrokes in the kernel to prevent interrupt-based attacks and \PrimeProbe attacks on the kernel. 
All keystrokes, including fake keystrokes, are carefully propagated through the shared library in order to hide any cache activity and thus to 
prevent \FlushReload attacks. Finally, we provide additional protection against \PrimeProbe for password input in user space programs. 
We show that attackers cannot distinguish fake keystrokes from real keystrokes anymore and we evaluate \KeyDrown on a commodity notebook as well as on two Android smartphones. 
We show that \KeyDrown eliminates any advantage an attacker can gain from using interrupt or cache side-channel information.
\end{abstract}

\maketitle

\section{Introduction}

Modern computer systems leak sensitive user information through side channels.
Among software-based side channels, information can leak, for example, from the system or from microarchitectural components such as the CPU cache~\cite{Ge2016} or the DRAM~\cite{Pessl2016}. 
Historically, side-channel attacks have exploited these information leaks to infer cryptographic secrets~\cite{Bernstein2005,Osvik2006,Yarom2014,Liu2015}, whereas more recent attacks even target keystroke timings and sensitive user input directly~\cite{Gruss2015Template,Oren2015,Pessl2016}.

In general, keystroke attacks aim to monitor when a keyboard input occurs, which either allows inferring user input directly or launching follow-up attacks~\cite{Song2001,Zhang2009}.
In particular, mobile devices may expose this information through sensor data, but practical mitigations are being deployed~\cite{Spreitzer2016systematic}.
Consequently, attackers are left with two different ways to obtain keystroke timings in a generic way. 
First, the \texttt{procfs} interface provides 
statistics for all interrupt sources, which allows monitoring the occurrence of keyboard interrupts. 
Second, microarchitectural attacks allow monitoring memory accesses with a granularity of single cache lines, and thus also allow recovering keystroke timings with a high accuracy.

Keystroke timing attacks are hard to mitigate, compared to side-channel attacks on cryptographic implementations.
Indeed, attacks on cryptographic implementations can be mitigated with changes in the algorithms, such as making execution paths independent of secret data.
On the contrary, user input travels a long way, from the hardware interrupt through the operating system and shared libraries up to the user space application. 
In order to detect a keystroke, an attacker just needs to probe a single spot in the keystroke path for activity. 

In the general case, keystrokes are non-repeatable low-frequency events, \ie if the attacker misses a keystroke, there is no way to repeat the measurement.
However, an attacker that explicitly targets a password field can record more timing traces when the user enters the password again.
While these traces have variations in timing, due to the variance of the typing behavior, it allows an attacker to combine multiple traces and to perform a more sophisticated attack. 
This makes attacks on password fields even harder to mitigate.

State-of-the art defense mechanisms only restrict access to the system interfaces providing interrupt statistics~\cite{Diao2016,Zhang2009}, and do not address all the layers involved in keystroke processing. 
Therefore, this does not prevent keystroke attacks at all. 
We first investigate interrupt-based attacks in a setting where the operating system does not provide any interface for interrupt statistics to the user. 
In such a restricted setting, we demonstrate two novel side-channel attacks to infer keystroke timings. 
The first attack uses the \texttt{rdtsc} instruction to determine the execution time of an interrupt service routine~(ISR), which is then used to determine whether or not the interrupt was caused by the keyboard.
The second attack uses \PrimeProbe on the kernel to determine when a keystroke is being processed in the kernel.

Based on these investigations and on state-of-the-art attacks, we identify three essential requirements for successful elimination of keystroke timing attacks on the entire software stack. In presence of the countermeasure:
\begin{compactenum}
	\item Any classifier based on a single-trace side-channel attack may not provide any advantage over a random classifier.
	\item The number of side-channel traces a classifier requires to detect all keystrokes correctly must be impractically high.
	\item The implementation of the countermeasure may not leak information about its activity or computations.
\end{compactenum}

Based on the identified requirements, we present \KeyDrown, a new defense mechanism against keystroke timing attacks exploiting microarchitectural side channels and interrupt side channels. 
\KeyDrown covers the entire software stack, from the interrupt source to the user space buffer storing the keystroke, both on x86 systems and on ARM devices.
We cover both the general case where an attacker can only obtain a single trace, and the case of password input where an attacker can obtain multiple traces.
\KeyDrown works in three layers:
\begin{compactenum}
	\item To mitigate interrupt-based attacks, \KeyDrown injects a large number of fake keyboard interrupts. 
	\PrimeProbe attacks on the kernel module are mitigated by unifying the control flow and data accesses of real and fake keystrokes such that there is no difference visible in the cache or in the execution time.
	\item To mitigate \FlushReload, and \PrimeProbe attacks on shared libraries, \KeyDrown runs through the same code path in the shared library for every fake and real keystroke.
	\item To mitigate \PrimeProbe attacks on password entry fields, \KeyDrown updates the widget buffer for every fake and real keystroke.
\end{compactenum}

We evaluate \KeyDrown on several state-of-the-art attacks as well as our two novel attacks. 
In all cases, \KeyDrown eliminates any advantage an attacker can gain from the side channels, \ie the attacker cannot deduce sensitive information from the side channel.

We provide a proof-of-concept implementation, which can be installed as a Debian package compatible with the latest long-term support release of Ubuntu. 
It runs on commodity operating systems with unmodified applications and unmodified compilers.
\KeyDrown is started automatically and is entirely transparent to the user, \ie requires no user interaction.
Although our countermeasure inherently executes more code than an unprotected system, it has no noticeable effect on keystroke latency.
Finally, we also define what \KeyDrown cannot protect against, such as word completion lookups or immediate forwarding of single keystrokes over the network.

\subheading{Contributions.} The contributions of this work are:
\begin{compactenum}
	\item We present two novel attack vectors to recover keystroke timings with fewer prerequisites than previous attacks.
	\item We identify three essential requirements for an effective countermeasure against keystroke attacks.
	\item We propose \KeyDrown, a multi-layered solution to mitigate keystroke timing attacks.\footnote{The \KeyDrown project is open-source and is available on GitHub: \mbox{\url{https://github.com/keydrown/keydrown}}.}
	\item We evaluate \KeyDrown and show that it eliminates all known
	attacks.
\end{compactenum}

\subheading{Outline.}
The remainder of the paper is organized as follows. 
In Section~\ref{sec:background}, we provide background information. 
In Section~\ref{sec:elimination}, we introduce our novel attacks and define requirements a defense mechanism has to provide to successfully mitigate attacks.
In Section~\ref{sec:design}, we describe the three layers of \KeyDrown.
In Section~\ref{sec:evaluation}, we demonstrate that \KeyDrown successfully mitigates keystroke timing attacks.
In Section~\ref{sec:limitations}, we discuss limitations and future work.
We conclude in Section~\ref{sec:conclusion}.

\section{Background}\label{sec:background}
In this section, we provide background information on interrupt handling as well as on software side channels that leak keystroke timing information. 

	\subsection{Linux Interrupt Handling}

Interrupt handling is one of the low-level tasks of an operating system and thus highly architecture and machine dependent. 
This section covers the general design of how interrupts and their handling within the Linux kernel work on both x86 PCs and ARMv7 smartphones. 

		\subsubsection{Interrupts on x86 and x86\_64}
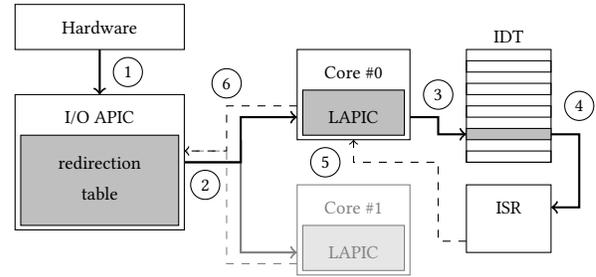
\begin{figure}
 \centering
\begin{tikzpicture}[xscale=0.75, yscale=0.6]
\footnotesize

\draw (0,4) rectangle +(3,1);
\node [text centered] at (1.5, 4.5) {Hardware};
\draw (0,0) rectangle +(3,3);
\node [text centered] at (1.5, 2.5) {I/O APIC};
\draw [fill=lightgray] (0.1,0.1) rectangle +(2.8,2);
\node [text centered] at (1.5, 1.5) {redirection};
\node [text centered] at (1.5, 0.8) {table};

\draw[->,thick] (1.5,4) -- (1.5,3);
\node [text centered] at (2, 3.5) {\circled 1};

\draw[->,thick,draw=gray] (3, 1.5) -| (4, 2) |- (5, -0.5);
\draw[->,thick] (3, 1.5) -| (4, 2) |- (5, 2.5);
\node [text centered] at (3.37, 1) {\circled 2};

\begin{scope}[shift={(0,0)}]
\draw (5,2) rectangle +(2,2);
\node [text centered] at (6, 3.5) {Core \#0};
\draw [fill=lightgray] (5.1,2.1) rectangle +(1.8,1);
\node [text centered] at (6, 2.5) {LAPIC};
\end{scope}

\begin{scope}[shift={(0,-3)}]
\draw[draw=gray] (5,2) rectangle +(2,2);
\node [text centered,color=gray] at (6, 3.5) {Core \#1};
\draw [fill=lightgray!40!white,draw=gray] (5.1,2.1) rectangle +(1.8,1);
\node [text centered,color=gray] at (6, 2.5) {LAPIC};
\end{scope}
\draw[->,dashed,draw=gray] (5, -0.75) -| (3.75, -0.75) |- (3.75, 1.75) |- (3, 1.75);

\draw (8,1.5) rectangle +(1.5,2.5);
\node [text centered] at (8.75, 4.3) {IDT};
\draw (8,1.5) rectangle +(1.5,0.25);
\draw (8,2.5) rectangle +(1.5,0.25);
\draw (8,3) rectangle +(1.5,0.25);
\draw (8,3.5) rectangle +(1.5,0.25);
\draw [fill=lightgray] (8,2) rectangle +(1.5,0.25);
\node [text centered] at (7.5, 3) {\circled 3};

\draw[->,thick] (7, 2.5) -| (7.5, 2.12) |- (8, 2.12);
\node [text centered] at (10, 2.5+0.25) {\circled 4};

\draw (8,-0.5) rectangle +(1.5,1.5);
\node [text centered] at (8.75, 0.5) {ISR};
\draw[->,thick] (9.5, 2.12) -| (10, 2.12) -- (10, 0.5) |- (9.5, 0.5);
\draw[->,dashed] (8, -0.25) -| (7.5, 0.5) |- (7.5, 1.5) -| (6, 1.5) |- (6, 2);
\node [text centered] at (5.5, 1.5) {\circled 5};
\draw[->,dashed] (5, 2.75) |- (3.75, 2.75) |- (3.75, 1.75) |- (3, 1.75);
\node [text centered] at (3.75, 3.25) {\circled 6};

\end{tikzpicture}
 \caption{Linux interrupt handling on x86.}
 \label{fig:linux-interrupt}
\end{figure}

Figure~\ref{fig:linux-interrupt} shows a high-level overview of interrupt handling on a dual-core x86 CPU. 
Interrupts are handled by the Advanced Programmable Interrupt Controller (APIC)~\cite{Intel_vol3}.
The APIC receives interrupts from different sources: locally and externally connected I/O devices, inter-processor interrupts, APIC internal interrupts, performance monitoring interrupts, and thermal sensor interrupts.
On multi-core systems, every CPU core has a local APIC (LAPIC) to handle interrupts. 
All LAPICs are connected to one or more I/O APICs which handle the actual hardware interrupts. 
The I/O APICs are part of the chipset and provide multi-core interrupt management by distributing the interrupts to the LAPICs as described in the ACPI system description tables~\cite{MicrosoftMADT2016}.

Interrupt-generating hardware, such as the keyboard, is connected to an I/O APIC pin (\circleds 1). 
The I/O APIC uses a redirection table to redirect hardware interrupts and the raised interrupt vector to the destination LAPIC (\circleds 2)~\cite{Intel_IOAPIC}.
In the case of multiple configured LAPICs for one interrupt, the I/O APIC chooses a CPU based on task priorities in a round-robin fashion~\cite{Bovet2005}.

The LAPIC receiving the interrupt vector fetches the corresponding entry from the Interrupt Descriptor Table (IDT) (\circleds 3) which is set up by the operating system. 
The IDT contains an offset to the Interrupt Service Routine (ISR) for every interrupt vector. 
The CPU saves the current CPU flags and jumps to the interrupt service routine (\circleds 4) which then handles the interrupt. 

After processing, the interrupt service routine acknowledges the interrupt by sending an end-of-interrupt (EOI) to the LAPIC (\circleds 5). 
It then returns using the \texttt{iret} instruction to restore the CPU flags and to enable interrupts again. 
The LAPIC forwards the EOI to the I/O APIC (\circleds 6) which then resets the interrupt line to enable the corresponding interrupt again. 

		\subsubsection{Interrupts on ARM}

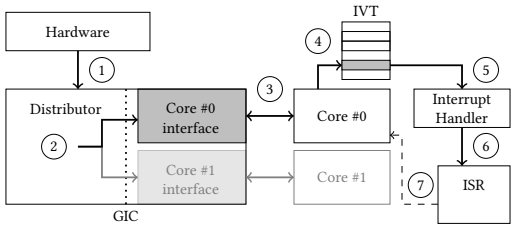
\begin{figure}
 \centering
 \resizebox{0.8\hsize}{!}{
\begin{tikzpicture}[xscale=0.75, yscale=0.6]
\footnotesize
\draw (0,4) rectangle +(3,1);
\node [text centered] at (1.5, 4.5) {Hardware};

\node [text centered] at (2.5, -0.3) {GIC};
\draw[dotted,thick] (2.5, 0) -- (2.5, 3);
\node [text centered] at (1.25, 2.5) {Distributor};
\draw [fill=lightgray!40!white,draw=lightgray] (2.75,0) rectangle +(2.25,1.4);
\draw [fill=lightgray] (2.75,1.6) rectangle +(2.25,1.4);

\node [text centered] at (3.85, 2.5) {Core \#0};
\node [text centered] at (3.85, 2) {interface};
\node [text centered,color=gray] at (3.85, 0.9) {Core \#1};
\node [text centered,color=gray] at (3.85, 0.4) {interface};
\draw (0,0) rectangle +(5,3);

\draw[->,thick,draw=gray] (1.5,1.5) -| (2, 0.7) |- (2.75,0.7);
\draw[->,thick] (1.5,1.5) -| (2, 2.2) |- (2.75,2.2);
\node [text centered] at (1, 1.5) {\circled 2};

\draw[->,thick] (1.5,4) -- (1.5,3);
\node [text centered] at (2, 3.5) {\circled 1};

\draw (6,1.6) rectangle +(2,1.4);
\node [text centered] at (7, 2.3) {Core \#0};
\draw [draw=gray] (6,0) rectangle +(2,1.4);
\node [text centered,color=gray] at (7, 0.7) {Core \#1};
\draw[<->,thick,gray] (5,0.7) -- (6,0.7);
\draw[<->,thick] (5,2.3) -- (6,2.3);
\node [text centered] at (5.5, 3) {\circled 3};

\begin{scope}[shift={(0,-0.25)}]
\draw (7,3.5) rectangle +(1,1.5);
\node [text centered] at (7.5, 5.3) {IVT};
\draw [fill=lightgray] (7,3.75) rectangle +(1,0.25);
\draw (7,4.25) rectangle +(1,0.25);
\draw (7,4.5) rectangle +(1,0.25);
\end{scope}

\draw[->,thick] (6.5,3) |- (7, 3.87-0.25);
\node [text centered] at (6.5, 4.25) {\circled 4};

\draw (8.5,1.75+0.25) rectangle +(2,1);
\node [text centered] at (10, 3.25+0.25) {\circled 5};
\node [text centered] at (9.5, 2.45+0.25) {Interrupt};
\node [text centered] at (9.5, 2+0.25) {Handler};
\draw[->,thick] (8,3.87-0.25) -| (9.5,2.75+0.25);

\draw (9,-0.5) rectangle +(1.5,1.5);
\node [text centered] at (9.75, 0.5) {ISR};
\draw[->,thick] (9.5,1.75+0.25) -- (9.5,1);
\node [text centered] at (10, 1.4+0.12) {\circled 6};

\draw[->,dashed] (9,0) -| (8.25,1.8) -- (8, 1.8);
\node [text centered] at (8.64, 0.5) {\circled 7};

\end{tikzpicture}}
 \caption{Linux interrupt handling on ARM.}
 \label{fig:linux-interrupt-arm}
\end{figure}

Figure~\ref{fig:linux-interrupt-arm} shows a high-level overview of interrupt handling on a dual-core ARMv7 CPU.
On ARM, interrupts are handled by the General Interrupt Controller (GIC). 
The GIC is divided into two parts, the distributor, and a CPU interface for every CPU core~\cite{ARM_GIC}. 
Every interrupt-generating device is connected to the distributor of the GIC (\circleds 1). 
The distributor (\circleds 2) schedules between CPU interfaces according to the interrupt's affinity mask.

When a CPU interface receives an interrupt, it signals it to the corresponding CPU core (\circleds 3). 
The core reads the interrupt number from the interrupt acknowledge register to acknowledge it. 
If the interrupt was sent to multiple CPU interfaces, all other CPU cores receive a spurious interrupt, as there is no more pending interrupt.

When receiving an interrupt, the CPU finishes executing the current instruction, switches to IRQ mode, and jumps to the IRQ entry of the Interrupt Vector Table (IVT) (\circleds 4).
The IVT contains exactly one instruction to jump to a handler function (\circleds 5).
In this handler function, the OS branches to the Interrupt Service Routine (ISR) corresponding to the interrupt number (\circleds 6). 

When the CPU is done servicing the interrupt, it writes the interrupt number to the End Of Interrupt register (\circleds 7) to signal that it is ready to receive this interrupt again~\cite{ARM_GIC_How}.

	\subsection{Microarchitectural Attacks}\label{sec:bg_ma}
CPU caches are a small and fast type of memory, buffering frequently used data to speed-up subsequent accesses.
There are typically three levels of caches in modern x86 CPUs, and two levels in modern ARM CPUs.
The last-level cache is typically shared across cores of the same CPU, which makes it a target for cross-core side-channel attacks.
On Intel x86 CPUs, the last-level cache is divided into one slice per core. 
The smallest unit managed by a cache is a cache line (typically \SI{64}{B}).
Modern caches are set-associative, \ie multiple cache lines are considered a set of equivalent storage locations.
A memory location maps to a cache set and slice based on the physical address~\cite{Maurice2015RAID,Yarom2015iacr,Inci2016CHES}.

		\paragraph{\FlushReload.}
\FlushReload~\cite{Yarom2014,Gullasch2011} is a technique that allows an attacker to monitor a victim's cache accesses at a granularity of a single cache line. 
The attacker flushes a cache line, lets the victim perform an operation, and then reloads and times the access to the cache line. 
A low timing indicates that the victim accessed the cache line. 
While very accurate, it can only be performed on shared memory, \ie shared libraries or binary code.
\FlushReload can neither be performed on dynamic buffers in a user program nor on code or data in the kernel.
Gruss~\etal\cite{Gruss2015Template} presented cache template attacks as a technique based on \FlushReload to automatically find and exploit cache-based leakage in programs.

		\paragraph{\PrimeProbe.}
\PrimeProbe~\cite{Percival2005,Osvik2006,Liu2015} is a technique that allows an attacker to monitor a victim's cache accesses at a granularity of a cache set. 
The attacker primes a cache set, \ie fills the cache set with its own cache lines. 
It then lets the victim perform an operation. 
Finally, it probes its own cache lines \ie measures the access time to them.
This technique does not require any shared memory between the attacker and the victim, but it is difficult due to the mapping between physical addresses and cache sets and slices. 
As \PrimeProbe only relies on measuring the latency of memory accesses, it can be performed on any part of the software stack. It is possible to perform \PrimeProbe on dynamically generated data~\cite{Lipp2016} as well as kernel memory~\cite{Osvik2006}.
Preventing \PrimeProbe attacks is difficult due to the huge attack surface and the fact that \PrimeProbe uses only innocuous operations such as memory accesses on legitimately allocated memory, as well as timing measurements.

		\paragraph{DRAMA}
Besides the cache, the DRAM design also introduces side channels~\cite{Pessl2016}, \ie timing differences caused by the DRAM row buffer. 
A DRAM bank contains a row buffer caching an entire DRAM row (\SI{8}{KB}). 
Requests to the currently active row are served from this buffer, resulting in a fast access, whereas other requests are significantly slower.
DRAM side-channel attacks do not require shared memory and work across CPUs of the same machine sharing a DRAM module.

	\subsection{Keystroke Timing Attacks}\label{sec:bg_kta}
	
		\paragraph{Keystrokes from Keystroke Timing.}
Keystroke timing attacks attempt to recover what was typed by the user by analyzing keystroke timing measurements.
These timings show characteristic patterns of the user, which depend on several factors such as keystroke sequences on the level of single letters, bigrams, syllables or words as well as keyboard layout and typing experience~\cite{Pinet2016}. Existing attacks train probabilistic classifiers like hidden Markov models or neural networks to infer known words or to reduce the password-guessing complexity~\cite{Song2001,Zhang2009,Simon2016don}.

Most keystroke timing attacks exploit the inter-keystroke timing, \ie the timing difference between two keystrokes, but according to Idrus~\etal\cite{Idrus2014} combinations of key press and key release events could also be exploited.
Pinet~\etal\cite{Pinet2016} report inter-keystroke interval values between \SI{160}{\milli\second} and \SI{200}{\milli\second} for skilled typists. 
Lee~\etal\cite{Lee2015} define the values depending on whether a text sequence was trained or entered for the first time, resulting in inter-keystroke intervals between \SI{125}{\milli\second} and \SI{215}{\milli\second} with a variance between \SI{43}{\milli\second} and \SI{106}{\milli\second}, again for trained and untrained text sequences. 

		\paragraph{Keystroke Timing from Software.}
A direct software side channel for keystroke timings is provided through OS interfaces~\cite{Song2001,Zhang2009}.
This includes instruction pointer and stack pointer information leaked through \texttt{/proc/stat}, interrupt statistics leaked through \texttt{/proc/interrupts}, and network packet statistics leaked through \texttt{/proc/net}~\cite{Zhang2009}.
As the instruction pointer and stack pointer information became too unpredictable, Jana and Shmatikov~\cite{Jana2012} showed that CPU usage yields much more reliable information on keystroke timings. 
Diao~\etal\cite{Diao2016} demonstrated high-precision keystroke timing attacks based on \texttt{/proc/interrupts}.
Vila~\etal\cite{Vila2017} recovered keystroke timings from timing differences caused by the event queue in the Chrome browser.

		\paragraph{Keystroke Timing from Microarchitectural Attacks.}
Gruss~\etal\cite{Gruss2015Template} demonstrated that \FlushReload allows distinguishing specific keys or groups of keys based on key-dependent data accesses in shared libraries.  
Ristenpart~\etal\cite{Ristenpart2009} demonstrated a keystroke timing attack using \PrimeProbe with a false-negative rate of \SI{5}{\percent} while measuring \SIx{0.3} false positive keystrokes per second. 
Pessl~\etal\cite{Pessl2016} showed that it is possible to use DRAM attacks to monitor keystrokes, \eg in the address bar of Firefox.
However, this attack only works if the target application performs a massive amount of memory accesses to thrash the cache reliably on its own.

\section{Keystroke Timing Attacks \& Defenses}\label{sec:elimination}
Due to the amount of code executed for every keystroke, there are many different side channels for keystroke timings.
In this section, we introduce our two novel attacks and compare them to state-of-the-art keystroke timing attack vectors, in order to understand the requirements for effective countermeasures. 
Finally, we derive three requirements for countermeasures to be effective against keystroke timing attacks.

The requirements are defined based on precision and recall of side-channel attacks.
The \textit{precision} is the fraction of true positive detected keystrokes in all detected keystrokes. 
If the precision is low, the side channel yields too many false positives to derive the correct keystroke timings.
The \textit{recall} is the fraction of true positive detected keystrokes in all real keystrokes. 
If the recall is low, \ie the side channel misses too many true positives, inter-keystroke timings are corrupted too.
A standard measure for accuracy is the \textit{\FScore}, \ie the geometric mean of precision and recall.
An \FScore of \SIx{1} describes a perfect side channel.
An \FScore of \SIx{0} describes that a side channel provides no information at all.

Note that there is only a limited number of keystroke time frames that can be reliably distinguished by an attacker, due to the number of keystrokes a user performs and the variance of inter-keystroke timing (\cf Section~\ref{sec:bg_kta}).
A keystroke timing attack providing nanosecond-accurate timestamps is actually only providing the binary information in which time frames a keystroke occurred.
Hence, we can compare side-channel-based classifiers to binary decision classifiers for these time frames.

An \textit{always-zero oracle} which never detects any event has an \FScore of \SIx{0}.
An \textit{always-one oracle} which ``detects'' an event in every possible time frame, \ie a large number of false positives, no false negatives, and no true negatives, is a channel which provides zero information.
Similarly, a \textit{random-guessing oracle}, which decides for every possible time frame whether it ``detects'' an event based on an apriori probability, also provides zero information.
For 8 keystrokes and 100 possible time frames per second, the \FScore for the always-one oracle is \SIx{\ValFscoreOneOracle} which is strictly better than the \FScore of the random-guessing oracle (\SIx{\ValFscoreRandOracle}).
An attacker relying on any side-channel-based classifier with a lower \FScore could achieve better results by simply using an always-one oracle, \ie in such a case it would not make sense to use the side-channel-based classifier in the first place.
In the remainder of the paper we assume that an attacker wants to find the real 8 keystrokes in 100 possible time frames per second.

	\subsection{Keystroke Timing Attack Surface}
Keystroke processing involves computations on all levels of the software stack. 
The keyboard interrupt is handled by one of the CPU cores, which interrupts the currently executed thread.
A significant amount of code is executed in the \textit{operating system kernel} and the keyboard driver until the preprocessed keystroke event is handed over to a user space \textit{shared library} that is part of the user interface.
The shared library distributes the keystroke event to all user interface elements listening for the event.
Finally, the shared library hands over the keyboard input to the active \textit{user space process} which further processes the input, \eg store a password character in a buffer.
This abundance of code and data that is executed and accessed upon a keystroke provides a multitude of possibilities to measure keystroke timings.

\setlength{\aboverulesep}{0pt}
\setlength{\belowrulesep}{0pt}
\begin{table}
	\centering
\caption{Comparison of keystroke timing attacks.}\label{tab:attacks}
\resizebox{\hsize}{!}{
\rowcolors{2}{white}{lightgray!40}
\begin{tabular}{p{3.2cm}ccc}
\toprule
	& Kernel & Shared library & User process \\ 
\midrule	
Interface-based	& \cmark{} \cite{Song2001,Zhang2009,Jana2012,Diao2016} & -- & -- \\ 
Timing-based & \cmark{} \textbf{new} & -- & -- \\ 
\FlushReload	& -- & \cmark{} \cite{Gruss2015Template}  & -- \\ 
\PrimeProbe on L1	& \cmark{} \cite{Ristenpart2009} & \cmark{} \cite{Ristenpart2009} & \cmark{} \cite{Ristenpart2009} \\ 
\PrimeProbe on LLC	& \cmark{} \textbf{new} & \cmark{} \textbf{new} & \cmark{} \textbf{new}  \\ 
DRAMA	& -- & -- & \cmark{} \cite{Pessl2016} \\ 
\bottomrule
\end{tabular} 
}
\end{table}

		\subsection{New Attack Vectors}\label{sec:new_attack_vectors}
Software side channels through \texttt{procfs} interfaces can be mitigated by restricting access to them~\cite{Diao2016,Zhang2009}.
However, such restrictions do not prevent keystroke timing attacks. 
We demonstrate two new attacks to infer keystroke timings: the first one exploits interrupt timings to detect keystrokes, and the second one relies on \PrimeProbe to attack a kernel module.
Table~\ref{tab:attacks} compares the novel attacks we describe in the following with the state-of-the-art attack vectors (\cf Section~\ref{sec:bg_kta})
in terms of attack techniques and the exploited attack surface.

			\paragraph{Low-Requirement Interrupt Timing Attack.}
We propose a new \emph{timing-based} attack that only requires unprivileged sand-boxed code execution on the targeted platform and an accurate timing source, \eg the \verb|rdtsc| instruction or a counter thread.
The basic idea is to monitor differences in the execution time of acquiring high-precision time stamps, \eg the \verb|rdtsc| instruction, as outlined in Algorithm~\ref{alg:record}.
While small differences between successive time stamps allow us to infer the CPU utilization, larger differences indicate that the measurement process was interrupted. 
In particular, I/O events like keyboard interrupts lead to clearly visible peaks in the execution time, due to the interaction of the keyboard ISR with hardware and the subsequent processing of keystrokes.

\begin{algorithm}[t]
	\SetKw{KwStep}{step}
	\SetKwInOut{Input}{input}\SetKwInOut{Output}{output}
	\Input{$\mathit{threshold}$, $N$}
	\Output{$\mathit{events[\,\,]}$, $\mathit{diff[\,\,]}$}
	\BlankLine
	$\mathit{tsc}[0] \leftarrow rdtsc()$\;
	\For{$i \in \gbr{1,\ldots,N}$}
	{
	  $\mathit{tsc}[i] \leftarrow rdtsc()$\;
	  \If{$\mathit{tsc}[i] - \mathit{tsc}[i-1] > \mathit{threshold}$} {
	    $\mathit{events}[$i$] \leftarrow \mathit{tsc}[i]$\;
   	    $\mathit{diff}[$i$] \leftarrow \mathit{tsc}[i] - \mathit{tsc}[i-1]$\;
	  }
	}
	\caption{Recording interrupt timing}\label{alg:record}
\end{algorithm}

\tikzstyle{every pin}=[fill=white, draw=black, font=\footnotesize]

\begin{figure}[t]
 \centering
 \begin{tikzpicture}[scale=0.95]
\begin{axis}[
xmin=100700000000,
xmax=105500000000,
xlabel={Time [cycles]},
ylabel={Delta [cycles]},
scaled y ticks=false,
ymin=0,
ymax=250000,
ytick={0,100000,200000},
yticklabels={0,100k,200k},
width=0.99\hsize,
height=3.5cm
]
    \addplot+[const plot,draw=none,fill=blue!40,opacity=0.9,text=blue,text opacity=1] table [x index=0,y expr=\thisrowno{1}*200000+3000,mark=none] {data/plainpwd.keys2.csv}
    [yshift=+5pt,font=\footnotesize]
            node[pos=0.12] {\texttt{g}\vphantom{X}}
            node[pos=0.35] {\texttt{\textbackslash n}\vphantom{X}}
            node[pos=0.64] {\texttt{d}\vphantom{X}}
            node[pos=0.73] {\texttt{i}\vphantom{X}}
            node[pos=0.87] {\texttt{1}\vphantom{X}}
            node[pos=0.975] {\texttt{3}\vphantom{X}}
    ;
    
    \addplot+[const plot,draw=none,fill=yellow,opacity=0.6,text=blue,text opacity=1] table [x index=0,y expr=\thisrowno{1}*200000+3000,mark=none] {data/plainpwd.keys1.csv}
    [yshift=+5pt,font=\footnotesize]
            node[pos=0.03] {\texttt{s}\vphantom{X}}
            node[pos=0.17] {\texttt{x}\vphantom{X}}
            node[pos=0.555] {\texttt{a}\vphantom{X}}
            node[pos=0.655] {\texttt{m}\vphantom{X}}
            node[pos=0.755] {\texttt{n}\vphantom{X}}
            node[pos=0.885] {\texttt{2}\vphantom{X}}
            node[pos=0.975] {\texttt{\textbackslash n}\vphantom{X}}
    ;

    \addplot+[blue,mark=*,mark size=0.5,mark options={fill=blue}] table [x index=0,y index=1] {data/plainpwd_full.csv};

\end{axis}
\end{tikzpicture} 
 \caption{Measured delta between continuous \texttt{rdtsc} calls while entering a password. Keystroke events interrupt the attacker and thus cause higher deltas.}
 \label{fig:inttiming_plain}
\end{figure}
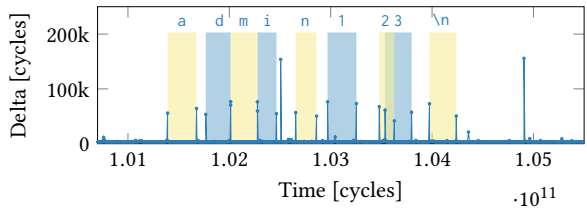

Figure~\ref{fig:inttiming_plain} illustrates these observations in a timing trace recorded while the user was typing a password. 
The bars indicate actual keystroke events, which almost perfectly match certain measurement points. 
Based on this plot, we can clearly distinguish keyboard interrupts (around \SIx{60000} cycles) from other interrupts. 
For example, rescheduling interrupts can be observed with a difference of about \SIx{155000} cycles. 
In this attack, we achieve a precision of \SIx{\ValPrecisionRDTSC} and a recall of \SIx{1}, resulting in an \FScore of \SIx{\ValFscoreRDTSC}, which means a significant advantage over an always-one oracle of +\SI{537.4}{\percent}. 

			\paragraph{\MultiPrimeProbe Attack on the Kernel.}
Our second attack relies on \PrimeProbe to attack the keyboard interrupt handler within the kernel. 
More specifically, we target the code in the keyboard interrupt handler that is executed each time a key is pressed. 
Thereby, keystroke events can be inferred by observing cache activity in the cache set used by the keyboard interrupt handler. 

To find the cache sets that are accessed by the keyboard interrupt handler, we first need to find the physical addresses where the code is located. 
We can either use the prefetch side-channel attack by Gruss~\etal\cite{Gruss2016CCS} or the TSX-based side channel by Jang~\etal\cite{Jang2016} to locate the code within the kernel.
Kernel Address-Space-Layout Randomization was not enabled by default until Ubuntu 16.10.
Thus, an attacker can also just use known physical addresses from an attacker-controlled system.

\begin{figure}[t]
 \centering
 \begin{tikzpicture}
\begin{axis}[mlineplot,
style={font=\footnotesize},
xlabel={Runtime [cycles]},
ylabel=Active cache sets,
width=1.0\hsize,
ymin=0,
xmin=0,
xmax=26744000000,
ymax=5.9,
height=3.5cm
]
\draw [draw=yellow, fill=yellow,opacity=0.6] (axis cs:1891409997,0.1) rectangle (axis cs:2670478278,5.8);
\draw [draw=yellow, fill=yellow,opacity=0.6] (axis cs:5870457411,0.1) rectangle (axis cs:6439866826,5.8);
\draw [draw=yellow, fill=yellow,opacity=0.6] (axis cs:10613485647,0.1) rectangle (axis cs:11392384006,5.8);
\draw [draw=yellow, fill=yellow,opacity=0.6] (axis cs:14378234523,0.1) rectangle (axis cs:15214610790,5.8);
\draw [draw=yellow, fill=yellow,opacity=0.6] (axis cs:18983596563,0.1) rectangle (axis cs:19857795736,5.8);

\addplot+[const plot,blue, no markers] table[x=Time,y=Value,col sep=comma] {data/prime_probe.csv};

\end{axis}
\end{tikzpicture}
 \caption{\MultiPrimeProbe attack on password input. Keystrokes cause higher activity in more cache sets.}
 \label{fig:kernel_pp}
\end{figure}
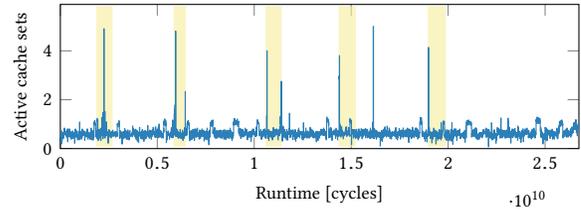

To reduce the influence of system noise, we developed a new form of \PrimeProbe attack called \MultiPrimeProbe. \MultiPrimeProbe combines the information from multiple simultaneous \PrimeProbe attacks on different addresses.
Figure~\ref{fig:kernel_pp} shows the result of such a \MultiPrimeProbe attack on the keyboard interrupt handler.
In a post-processing step we smoothed the \MultiPrimeProbe trace with a \SI{500}{\micro\second} sliding window.
The keystroke events cause higher activity in the targeted cache sets and thus produce clearly recognizable peaks for every key event.
Despite doubts that such an attack can be mounted~\cite{Gruss2016Flush}, our attack is the first highly accurate keystroke timing attack based on \PrimeProbe on the last-level cache. 
More specifically, we achieve a precision of \SIx{\ValPrecisionPP} and a recall of \SIx{\ValRecallPP}, resulting in an \FScore of \SIx{\ValFscorePP}, which is significantly better than state-of-the-art \PrimeProbe attacks. 

	\subsection{Requirements for Effective Elimination of Keystroke Timing Attacks}\label{sec:requirements}
As demonstrated in the previous section, we are able to craft new attacks with fewer requirements than state-of-the-art attacks. 
Hence, countermeasures against keystroke timing attacks must be designed in a generic way, in all affected layers of the software stack, covering known and unknown attacks. 

\paragraph{Attack Model.}
We assume an attacker can run an unprivileged program on the target machine. The attacker is thus able to continuously run a side-channel attack and obtain traces for all user input. We assume the countermeasure was already installed when the attacker gained unprivileged access to the machine. Consequently, the attacker cannot obtain keystroke timing templates and thus cannot perform a template attack.

We assume an attacker can generally obtain only a single trace for any user input sequence, but multiple traces for password input. 
In contrast to side-channel attacks on algorithms, which can be repeated multiple times, user input sequences are generally not (automatically) repeatable and thus an attacker cannot obtain multiple traces. An exception are phrases that are repeatedly entered in the same way, such as login credentials and especially passwords. A countermeasure must address both cases.

To effectively eliminate keystroke timing attacks, we identify the three following requirements a countermeasure must fulfill.

\paragraph{\ReqOne: Minimize Side Channel Accuracy.}
As user input sequences are in general not (automatically) repeatable, keystroke timing attacks require a high precision and high recall to succeed.
To be effective, a countermeasure must reduce the \FScore enough so that the attacker does not gain any advantage from using the side channel over an always-one oracle. 
More specifically, the \FScore of the side-channel based classifier may not be above the \FScore of the always-one oracle (\SIx{\ValFscoreOneOracle}). 
Ristenpart~\etal\cite{Ristenpart2009} reported a false-negative rate of \SI{5}{\percent} with \SIx{0.3} false positives per second.
At an average typing speed for a skilled typist of \SIx{8} keystrokes per second~\cite{Pinet2016}, the \FScore is thus \SIx{\ValFscoreRistenpart}, which is an advantage over an always-one oracle of +\SI{545.3}{\percent}.
Gruss~\etal\cite{Gruss2015Template,Gruss2016Flush} reported false-negative rates $\leq$ \SI{8}{\percent} with no false positives, resulting in an \FScore of $>$ \SIx{\ValFscoreGruss}, which is an advantage over an always-one oracle of +\SI{546.9}{\percent}. 
Thus, we assume a countermeasure is effective if it reduces the \FScore of 
side channels significantly, such that using the side channel gives an advantage over an always-one oracle of $\leq$\SI{0.0}{\percent}.

\paragraph{\ReqTwo: Reduction of Statistical Characteristics in Password Input.}
In case of a password input, we assume that an attacker can combine information from multiple traces, \ie exploit statistical characteristics.
A countermeasure is effective if the attacker requires an impractical number of traces to reach the \FScore of state-of-the-art attacks, \ie higher than $0.95$. 

Specifically, if the side-channel attack requires more traces than can be practically obtained, we consider the side-channel attack not practical.
Studies~\cite{Gaw2006,Shay2010,CSID2012,Das2014,Wash2016} estimate that most users have 1--5 different passwords and enter \SIx{5} passwords per day on average.
It is also estimated that $56\%$ of users change their password at least once every 6 months. 
Thus, even if we assume that we attack a user with a single password that is entered \SIx{5} times per day, the expected number of measurement traces that an attacker is able to gather after 6 months is 913. 
Assuming that attackers might come up with new side-channel attacks, a generous security margin must be applied.
We consider a countermeasure effective if it requires more than \SIx{\ValMinTraces} traces, \ie traces for a whole year, to reach an \FScore of \SIx{0.95}. 

\paragraph{\ReqThree: Implementation Security.}
\ReqOne and \ReqTwo define how the countermeasure must be designed to be effective. 
However, the implementation itself can indirectly violate \ReqOne or \ReqTwo by leaking side-channel information on computations of the countermeasure itself. Consequently, an attacker may be able to filter the true positive keystrokes.
We thus require that the countermeasure may not have distinguishable code paths or data access patterns to guarantee that it is free from leakage.

If the implementation does not leak by itself, an attacker is only left with the low \FScore{s} from \ReqOne and \ReqTwo. If all requirements are met, classical password recovery attacks like brute force and more sophisticated attacks using Markov $n$-grams~\cite{Ma2014, Narayanan2005}, probabilistic context-free grammars (PCFG)~\cite{Weir2009, Veras2014}, or neural networks~\cite{Melicher2016}, are more practical than a side-channel attack in the presence of the countermeasure.

In the following section we describe the design of a countermeasure that fulfills all three requirements.

\section{\KeyDrown Multi-layer Design}\label{sec:design}

\begin{figure}[t]
 \centering
 \input{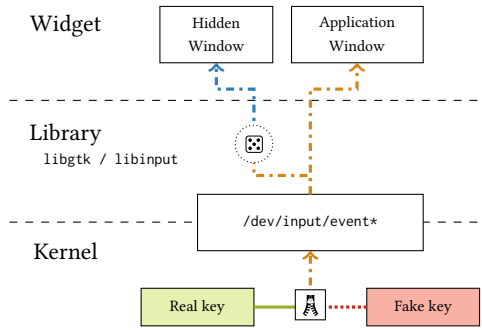}
 \caption{Multi-layered design of \KeyDrown.}
 \label{fig:keydrown_layer}
\end{figure}

We designed \KeyDrown as a multi-layered countermeasure.\footnote{The \KeyDrown project is open-source and is available on GitHub: \mbox{\url{https://github.com/keydrown/keydrown}}.}
Each layer builds up on the layer beneath and adds additional protection. 
Figure~\ref{fig:keydrown_layer} shows how the layers are connected to each other. 
The first layer implements a protection mechanism against interrupt-based attacks and timing-based attacks by artificially injecting interrupts.
The injected interrupts mimic user behavior to hide the real interrupt within a multitude of fake interrupts.  
All keystrokes, \ie real keystrokes and fake keystrokes, are passed to the library in a way which is indistinguishable for an attacker. 
The second layer protects the library handling the user input against \FlushReload attacks, including cache template attacks, and \PrimeProbe attacks. 
For every keystroke event received from the kernel, a random keystroke is sent to a hidden window. The library cannot distinguish between real and fake keystrokes and thus both have the same execution path.
In the third layer, the actual password entry field is protected against \PrimeProbe attacks by accessing the underlying buffer whenever a real or a fake keystroke is received. 

	\subsection{First Layer}
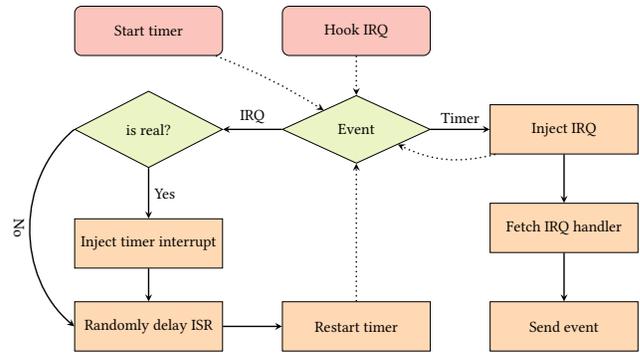
\begin{figure}[t]
 \tikzstyle{startstop} = [rectangle, rounded corners, minimum width=3cm, minimum height=1cm,text centered, draw=black, fill=red!30]
\tikzstyle{io} = [trapezium, trapezium left angle=70, trapezium right angle=110, minimum width=3cm, minimum height=1cm, text centered, draw=black, fill=blue!30]
\tikzstyle{process} = [rectangle, minimum width=3cm, minimum height=1cm, text centered, draw=black, fill=orange!30]
\tikzstyle{decision} = [diamond, minimum width=3cm, minimum height=1cm, text centered, draw=black, fill=green!30]
\tikzstyle{arrow} = [thick,->,>=stealth]
\usetikzlibrary{shapes.geometric, arrows}

\resizebox{\hsize}{!}{
\begin{tikzpicture}[node distance=2cm,scale=1,transform shape]

\node (hookirq) [startstop] {Hook IRQ};
\node (starttimer) [startstop, left of=hookirq, xshift=-7em] {Start timer};
\node (event) [decision, below of=hookirq] {Event};
\node (isreal) [decision, left of=event, xshift=-7em] {is real?};
\node (injectirq) [process, right of=event, xshift=7em] {Inject IRQ};
\node (injecttimer) [process, below of=isreal,yshift=-1em] {Inject timer interrupt};
\node (delay) [process, below of=injecttimer,yshift=1em] {Randomly delay ISR};
\node (prefetch) [process, below of=injectirq] {Fetch IRQ handler};
\node (send) [process, below of=prefetch] {Send event};
\node (restart) [process, below of=event,yshift=-2cm] {Restart timer};

\draw [arrow,dotted] (hookirq) -- (event);
\draw [arrow,dotted] (restart) -- (event);
\draw [arrow] (injectirq) -- (prefetch);
\draw [arrow] (prefetch) -- (send);
\draw [arrow] (injecttimer) -- (delay);
\draw [arrow] (delay) -- (restart);

\draw [arrow] (event) -- node[anchor=south] {Timer} (injectirq);
\draw [arrow] (event) -- node[anchor=south] {IRQ} (isreal);
\draw [arrow] (isreal) -- node[anchor=west] {Yes} (injecttimer);
\draw [bend right=50, arrow] (isreal.west) to node[below,sloped] {No}  (delay.west);
\draw [bend left=5, arrow,dotted] (starttimer) to (event);
\draw [bend left=20, arrow,dotted] (injectirq) to (event);

\end{tikzpicture}
}
 \caption{General flowchart of the kernel module.}
 \label{fig:kernel-flowchart}
\end{figure}

\paragraph*{Basic Concept.}
Figure~\ref{fig:kernel-flowchart} shows the program flow for the kernel part of \KeyDrown for both x86 and ARM.
We use recurrent timer interrupts with random delays to inject fake keystrokes.
Note that this leads to a uniform random distribution of keystrokes over time.
The kernel module handles two types of events, namely hardware interrupts from the input device, and the recurrent timer interrupt.
If the kernel module receives a timer interrupt, it injects a keyboard interrupt.
If it receives a keyboard interrupt, it injects a timer interrupt. Thus, for real and fake keystrokes both interrupts occur.
To minimize the effect of the real keyboard interrupt on the interrupt density, the next recurrent timer interrupt is rescheduled with a random delay. 
This guarantees that overall, the keystroke interrupt density remains uniform and real keystrokes cannot be distinguished from fake keystrokes.

For the fake keystrokes, the kernel uses a typically unused key value. 
The kernel does not have varying code paths and data accesses based on the key value, hence, the same code is executed for both real and fake keystrokes.
In both cases, the keystroke handler is delayed by a small random delay to hide timing differences from interrupt runtimes.
Finally, all keystrokes are passed to the library through the same data structures (\cf Figure~\ref{fig:keydrown_layer}).
Consequently, the attacker cannot use a \PrimeProbe or \MultiPrimeProbe attack on the kernel to distinguish real and fake keystrokes.

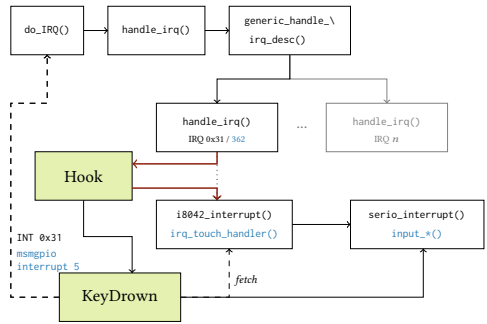
\begin{figure}[t]
 \centering
 \resizebox{0.75\hsize}{!}{\begin{tikzpicture}[scale=0.9]
\draw (0, 0) rectangle +(1.5, 1);
\node [text centered] at (0.75, 0.5) {\scriptsize \texttt{do\_IRQ()}};
\draw[->] (1.5, 0.5) -- (2, 0.5);

\draw (2, 0) rectangle +(2, 1);
\node [text centered] at (3, 0.5) {\scriptsize \texttt{handle\_irq()}};
\draw[->] (4, 0.5) -- (4.5, 0.5);

\draw (4.5, 0) rectangle +(2.5, 1);
\node [text centered] at (5.75, 0.7) {\scriptsize \texttt{generic\_handle\_\textbackslash}};
\node [text centered] at (5.5, 0.3) {\scriptsize \texttt{irq\_desc()}};
\draw[->,draw=gray] (5.75, 0) |- (7.75, -0.5) -- (7.75, -1);
\draw[->] (5.75, 0) |- (4.25, -0.5) -- (4.25, -1);

\draw (3, -2) rectangle +(2.5, 1);
\node [text centered] at (4.25, -1.4) {\scriptsize \texttt{handle\_irq()}};
\node [text centered] at (4.25, -1.75) {\tiny {IRQ 0x31 / \textcolor{blue}{362}}};

\begin{scope}[shift={(3.5, 0)}]
\draw [color=gray] (3, -2) rectangle +(2.5, 1);
\node [text centered,color=gray] at (4.25, -1.4) {\scriptsize \texttt{handle\_irq()}};
\node [text centered,color=gray] at (4.25, -1.75) {\tiny {IRQ $n$}};
\node [text centered,color=gray] at (2.5, -1.5) {...};
\end{scope}

\draw (3, -4) rectangle +(2.8, 1);
\node [text centered] at (4.4, -3.3) {\scriptsize \texttt{i8042\_interrupt()}};
\node [text centered,color=blue] at (4.4, -3.7) {\scriptsize \texttt{irq\_touch\_handler()}};
\draw[->,dotted] (4.25, -2) -- (4.25, -3);

\draw [fill=green!40!white] (0.5, -3) rectangle +(2, 1);
\node [text centered] at (1.5, -2.5) {Hook};
\draw[->,draw=red!60!black,thick] (4.25, -2) |- (2.5, -2.25);
\draw[->,draw=red!60!black,thick] (2.5, -2.75) -| (4.25, -3);

\draw (7, -4) rectangle +(2.75, 1);
\node [text centered] at (8.37, -3.3) {\scriptsize \texttt{serio\_interrupt()}};
\node [text centered,color=blue] at (8.37, -3.7) {\scriptsize \texttt{input\_*()}};
\draw[->] (5.8, -3.5) -- (7, -3.5);

\draw [fill=green!40!white] (1, -5.5) rectangle +(2.5, 1);
\node [text centered] at (2.25, -5) {KeyDrown};
\draw[->] (1.5, -3) |- (2.5, -3.75) -- (2.5, -4.5);
\draw[->] (3.5, -5) -|  (8.5, -4);
\draw[->,dashed] (3.5, -5) -| node[anchor=west,yshift=1em] {\scriptsize \textit {fetch}} (4.5, -4);
\draw[->,dashed] (1, -5) -| node[anchor=west,yshift=3.5em] {\scriptsize \texttt {INT 0x31}} (0, -1) -| (0.75, 0);
\node[anchor=west,yshift=2.5em,color=blue] at (0, -5) {\scriptsize \texttt {msmgpio}};
\draw[->,dashed] (1, -5) -| node[anchor=west,yshift=1.75em,color=blue] {\scriptsize \texttt {interrupt 5}} (0, -1) -| (0.75, 0);

\end{tikzpicture}}
 \caption{Linux kernel module design for x86 and the Snapdragon SoC. Snapdragon specific functions are marked in blue.}
 \label{fig:kernel-module}
\end{figure}

\paragraph*{Implementation Details.}
The first layer of \KeyDrown is implemented as a Linux kernel module that aims to prevent interrupt-based attacks on keystrokes. 
We do not require a custom kernel or any patches to the Linux kernel itself. 
All functionality is implemented in one generic kernel module that can be loaded into any Linux kernel from version 3.4 to 4.10, the newest release at the time of writing. 
The interrupt hardware and handling mechanism is compatible with all personal computers, thus there is no further limitation on PC hardware or Linux distributions. 

Figure~\ref{fig:kernel-module} shows the implementation details of the \KeyDrown kernel module.
The recurrent timer interrupts are implemented using the Linux platform-independent high-resolution timer API~\cite{LWN_hrtimer}.
On Linux, a driver can register an interrupt handler for a specific interrupt which is called whenever the CPU receives the interrupt. 
The interrupt service routine \texttt{do\_IRQ} calls the general \texttt{handle\_irq} function which subsequently calls \texttt{generic\_handle\_irq\_desc} to execute the correct handler for every interrupt.
To receive all hardware interrupts, we change the input device's interrupt handler to a function within our kernel module. 
Afterwards we forward the interrupt to the actual input device driver (\ie \texttt{i8042\_interrupt} on x86, and \texttt{irq\_touch\_handler} on the \Nexus).
Every time the kernel receives a recurrent timer interrupt or a real hardware interrupt, we restart the recurrent timer with a new random delay to maintain the uniform random distribution over time.

The kernel module triggers a hardware interrupt for every recurrent timer interrupt.
On x86, we can simply execute the \texttt{int} assembly instruction with the corresponding interrupt number. 
This spurious keyboard interrupt travels up until the point where the keyboard driver tries to read the scancode from the hardware.
As the driver does not execute the entire \texttt{i8042\_interrupt} function for spurious interrupts, we access the remaining function to fetch it into the cache as if it was executed.
In contrast, for real keys we access the code that injects the keys to fetch it into the cache as if it was executed.
From an attacker's point of view, there is no difference in cache activity between a data fetch and a code fetch, \ie a \PrimeProbe attack cannot determine the difference.

We inject a scancode of a typically unused key, such as F16 or a Windows multimedia key using the standard \texttt{serio\_interrupt} interface. 
Thus, from this point on the only difference between real and fake keystrokes is the scancode. 
Finally, all scancodes are sent to the upper software layers and run through the same execution path.

On the ARM platform, hardware interrupts and device drivers are hardware dependent. 
We decided to implement our proof-of-concept on the widespread Qualcomm Snapdragon Mobile Station Modem (MSM) SoC~\cite{QualcommSnapdragon2017}. 

ARM processors generally do not provide an assembly instruction to generate arbitrary interrupts from supervisor mode. 
Instead, we have to communicate with the interrupt controller directly. 
The Snapdragon MSM SoC implements its own intermediate I/O interrupt controller. 
All interrupt generating hardware elements are connected to this interrupt controller and not directly to the GIC. 
Therefore, if we want to inject an interrupt, we write the interrupt state of the touchscreen interrupt via memory mapped I/O registers to the MSM I/O interrupt controller.
The remaining execution path is analogous to the x86 module. 
When the driver aborts due to a spurious interrupt, we fetch the \texttt{irq\_touch\_handler} to produce the same cache footprint as if it is executed.
We inject an out-of-bounds touch event using the \texttt{input\_event}, \texttt{input\_report\_abs}, and \texttt{input\_sync} functions, which is then handed to the upper layers.

	\subsection{Second Layer}
\paragraph*{Basic Concept.}
The second layer countermeasure ensures that the control flow within the key-handling library is exactly the same for both real and injected keystrokes.
The fundamental idea of the second layer is that real and injected keystrokes should have the same code paths and data accesses in the library. 
We rely on the events injected in the first layer to propagate them further through the key-handling library.
The injected keys sent by our first layer are valid, but typically unused keys, thus they travel all the way up to the user space and are received by the userspace application. 
However, these unused keys might not have the exact same path within the library. 

Gruss~\etal showed that an attacker can build cache template attacks based on \FlushReload~\cite{Gruss2015Template} to detect keystrokes and even distinguish groups of keys. 
This cache leakage can also be measured with \MultiPrimeProbe.
Both attacks exploit the cache activity of certain functions that are only called if a keystroke is handled, \ie varying execution paths and access patterns~\cite{Gruss2015Template}.
We mitigate these attacks by duplicating every key event (\cf Figure~\ref{fig:keydrown_layer}) running through multiple execution paths and access sequences simultaneously. 
The key value of the duplicated key event is replaced by a random key value and the key event is sent to a hidden window.
Hence, the two key events, the real and the duplicated one, are processed simultaneously by the remainder of the library and the two applications.
This introduces a significant amount of noise on cache template attacks on the library layer.

The real key event at this point may still be a fake keystroke from the kernel.
However, we duplicate the key event in order to trigger key value processing and key drawing in the library and the hidden window for both fake and real keystrokes.
Consequently, we cannot distinguish real and fake keystrokes on the library layer using a side channel anymore.

\paragraph*{Implementation Details.}
One of the most popular user interface libraries for Linux is \GTKplus~\cite{Xdocu2014}.
The \GTKplus library handles the user input for many desktop environments and is thus included in most Linux distributions~\cite{GTKplus2016}. 
As we cannot hide cache activity, we have to generate artificial cache activity for the same cache lines that are active when handling real user inputs. 

The kernel provides all events, such as keyboard inputs, through the \texttt{/dev/input/event*} pseudo-files to the user space. 
The X Window System uses these files to provide all events to the \GTKplus event queue. 

On x86, the second layer is a standalone \GTKplus application. 
On system startup, we create a hidden window containing a text field. 
The application uses \texttt{poll} to listen to the \texttt{/dev/input/event*} interface to get notified whenever a keyboard event occurs. 
This allows \KeyDrown to have a very low performance overhead, as the application is not using CPU time as long as it is waiting inside the \texttt{poll} function. 
Whenever we receive a keystroke event from the kernel, we create an additional \GTKplus keystroke event with a random key that is associated with the text field of the hidden window. 
For every keystroke --- regardless of whether it is a printable character or not --- that comes from the kernel, the same path is active within the library. 
Thus, an attacker cannot distinguish an injected keystroke from a real keystroke anymore. 

The second layer has no knowledge of an event's source. 
Thus, it cannot violate \ReqThree, as the information whether a keystroke is real or injected is not present within the second layer. 

On Android, the handling of input events is considerably simpler. 
The injected events travel directly to the foreground application without going to any non-Android library. 
Thus, all events have exactly the same execution path and it is only necessary to drop our fake event immediately before the registered touch event handler is called.
To not leak any information through the non-executed touch handler, we access the cache lines in the same way as if the touch handler was executed.

	\subsection{Third Layer}
\paragraph*{Basic Concept.}
While the first layer protects against interrupt-based attacks and the second layer prevents attacks on the library handling the user inputs, the buffer that stores the actual secret is not protected and can still be monitored using a \PrimeProbe attack. 
The fake keystrokes sent by the kernel are unused key codes, which do not have any effect on the user interface element or the corresponding buffer.
We mitigate cache attacks on this layer by generating cache activity on the cache lines that are used when the buffer is processed for any key code received from the kernel.
More specifically, we access the buffer every time the library receives a keystroke event from the kernel.
This ensures that the buffer is cached for both real and fake keystrokes. 

An attacker who mounts a \FlushReload attack against the library, or a \PrimeProbe attack directly on the buffer, sees cache activity for both real and injected events. 
This is also the case for cache template attacks, as the injected events induce a significant amount of noise in both the profiling and the exploitation phase.
Therefore, the third layer protects against attacks that are mounted against the Android keyboard as shown by Lipp~\etal\cite{Lipp2016}, or \MultiPrimeProbe attacks directly on the input field buffer (cf.~Section~\ref{sec:new_attack_vectors}). 

\paragraph*{Implementation Details.}
In \GTKplus the \texttt{GtkEntry} widget is used as a single-line text and password entry field. 
By setting its \textit{visibility} flag, entered characters are replaced by a symbol and, thus, hidden from the viewer. 
The \texttt{GtkEntry} widget implements the \texttt{GtkEditable} interface that describes a text-editing widget. 

Implementing the countermeasure directly in the \GTKplus library would require to rebuild the library and all of its dependencies. 
As this is highly impractical, we chose a different approach: \texttt{LD\_PRELOAD} allows listing shared objects that are loaded before other shared objects on execution of the program~\cite{LD_PRELOAD}. 
By using this environment variable, we can overwrite the \texttt{gtk\_entry\_new} function that is called when a new object of \texttt{GtkEntry} should be created. 
In our own implementation, we register a key press event handler for the new entry field. 
This event handler is called on both real and injected keys and accesses the underlying buffer. 

On Android, the basic concept is the same. 
It is, however, implemented as part of the keyboard and not the library. 
The keyboard relies on the \texttt{inotifyd} command to detect touch events provided by the kernel. 
If a password entry field is focused, the keyboard accesses the password entry buffer on every touch event by calling the key handling function with a dummy key. 
This ensures that both the buffer as well as the keyboard's key handling functions are active for every event.

\section{Evaluation}\label{sec:evaluation}
We evaluate \KeyDrown with respect to the requirements \ReqOne, \ReqTwo, \ReqThree as well as discuss the performance of our implementation. 
We evaluate the x86 version of \KeyDrown on a Lenovo ThinkPad T460s with an Intel Core i5-6200U and the ARM version on both an LG \Nexus (ARMv7) and a \OnePlus (ARMv8). 
A large comparison table can be found in Appendix~\ref{app:comparison}.
As the results are very similar for all architectures we provide the results for the LG \Nexus (ARMv7) in Appendix~\ref{app:nexus} and for the \OnePlus (ARMv8) in Appendix~\ref{app:op3}. 
We evaluate four different side channels with and without \KeyDrown: \texttt{procfs}, \texttt{rdtsc}, \FlushReload (including cache template attacks), and \PrimeProbe on the last-level cache.
We also discuss \PrimeProbe attacks on the L1 cache and DRAMA side-channel attacks.

To evaluate \KeyDrown, we chose a uniform key-injection interval $[\SI{0}{\milli\second},\SI{20}{\milli\second}]$ resulting in one event (either a real or an injected keystroke) every \SI{10}{\milli\second} on average.
Thus, on average we expect \SIx{100} events per second.

As described in Section~\ref{sec:elimination}, we compare our results to an always-one oracle and a random-guessing oracle.
A random-guessing oracle, which chooses randomly---without any information---for every \SI{10}{\milli\second} interval whether there was a keystroke based on an apriori probability, would achieve an \FScore of \SIx{\ValFscoreRandOracle}.
The always-one oracle performs slightly better, as it has a higher true positive rate of \SI{100}{\percent}, but it also has a false positive rate of \SI{100}{\percent}, \ie the oracle neither uses nor provides any information.
The \FScore of the always-one oracle is \SIx{\ValFscoreOneOracle} and thus, higher than the \FScore of a random-guessing oracle.
If a side channel yields an \FScore of this value or below, the attacker gains no advantage over the always-one oracle from this side channel.

For all evaluated attacks, we provide the precision of the attack with and without \KeyDrown, based on the best threshold distinguisher we can find.
\KeyDrown does not influence the recall, as it does not reduce the number of true positives and it also does not increase the number of real keystrokes.
However, we provide the recall for all attacks with a recall below \num{1}. 
The harmonic mean of precision and recall---the \FScore---gives an indication how well the countermeasure works. 
We provide the advantage over the always-one oracle as a direct indicator on whether it makes sense to use the side channel or not.

	\subsection{Requirement \ReqOne}

We evaluate \KeyDrown with respect to \ReqOne, the elimination of single-trace attacks.
\ReqOne defines that a side channel may not provide any advantage over an always-one oracle, \ie the advantage measured in the \FScore must be $\leq$\SI{0.0}{\percent}.
We show that \KeyDrown fulfills this requirement by mounting state-of-the-art attacks with and without \KeyDrown.
Table~\ref{tab:fscore} summarizes the \FScore{s} for all attacks with and without \KeyDrown. 
In all cases, \KeyDrown eliminates any advantage that can be gained from the side channel, when considering single-trace attacks only.
In some cases, the numerous false positives and false negatives lead to an even worse \FScore.

\setlength{\aboverulesep}{0pt}
\setlength{\belowrulesep}{0pt}
\begin{table}
	\centering
\caption{\FScore without and with \KeyDrown and advantage over always-one oracle for state-of-the-art attacks. \KeyDrown eliminates any side-channel advantage.}\label{tab:fscore}
\resizebox{\hsize}{!}{
\rowcolors{2}{white}{lightgray!40}
\begin{tabular}{p{2.5cm}cccc}
\toprule
Side Channel& no \KeyDrown & ($\Delta$ always-one) & \KeyDrown & ($\Delta$ always-one) \\ 
\midrule	
\texttt{procfs} & \SIx{\ValFscoreProc} & (+\SI{575.0}{\percent}) & \SIx{\ValFscoreProcKD} & (+\SI{0.0}{\percent}) \\ 
\texttt{rdtsc} & \SIx{\ValFscoreRDTSC} & (+\SI{537.4}{\percent}) & \SIx{\ValFscoreRDTSCKD} & (\SI{-3.8}{\percent}) \\ 
\FlushReload	& \SIx{\ValFscoreFR} & (+\SI{569.3}{\percent}) & \SIx{\ValFscoreFRKD} & (\SI{-40.2}{\percent}) \\ 
LLC \PrimeProbe	& \SIx{\ValFscorePP} & (+\SI{440.0}{\percent}) & \SIx{\ValFscorePPKD} & (\SI{-27.7}{\percent}) \\ 
\bottomrule
\end{tabular} 
}
\end{table}

\begin{figure}[t!]
    \centering
    \begin{subfigure}[t]{0.99\hsize}
      \centering
      \begin{tikzpicture}
\begin{axis}[
mlineplot,
style={font=\footnotesize},
xlabel={Runtime [cycles]},
ylabel={Latency [cycles]},
width=1.0\hsize,
xmin=0,
ymax=630,
ymin=0,
height=3.5cm
]
\addplot+[blue, no markers] table[x=Time,y=Value,col sep=comma] {data/layer2_flush_reload_without_keydrown.csv};
\addplot+[only marks,mark options={draw=green,fill=green},mark=*] table[x=Time,y=Real,col sep=comma] {data/layer2_flush_reload_without_keydrown.csv};
\addplot+[only marks, mark options={draw=red,fill=red},mark=*] table[x=Time,y=Fake,col sep=comma] {data/layer2_flush_reload_without_keydrown.csv};

\end{axis}
\end{tikzpicture}
      \caption{Without \KeyDrown.}
      \label{fig:layer2_fr_kd_disabled}
    \end{subfigure}%

    \begin{subfigure}[t]{0.99\hsize}
      \centering
      \begin{tikzpicture}
\begin{axis}[
mlineplot,
style={font=\footnotesize},
xlabel={Runtime [cycles]},
ylabel={Latency [cycles]},
width=1.0\hsize,
xmin=0,
ymax=630,
ymin=0,
height=3.5cm
]
\addplot+[blue, no markers] table[x=Time,y=Value,col sep=comma] {data/layer2_flush_reload_with_keydrown.csv};
\addplot+[only marks,mark options={draw=green,fill=green}, mark=*] table[x=Time,y=Real,col sep=comma] {data/layer2_flush_reload_with_keydrown.csv};
\addplot+[only marks, mark options={draw=red,fill=red},mark=triangle*] table[x=Time,y=Fake,col sep=comma] {data/layer2_flush_reload_with_keydrown.csv};

\end{axis}
\end{tikzpicture}
      \caption{With \KeyDrown.}
      \label{fig:layer2_fr_kd}
    \end{subfigure}%

    \caption{\FlushReload attack on address \texttt{0x381c0} of
      \texttt{libgdk-3.so}. (a) The attack allows to clearly detect every single
      keystroke (\RealMarker). (b) With \KeyDrown, the attacker
      measures cache hits on injected keystrokes (\FakeMarker) as well as on real events (\RealMarker) and cannot distinguish
    between them.}
    \label{fig:layer2_fr_gdk}
\end{figure}
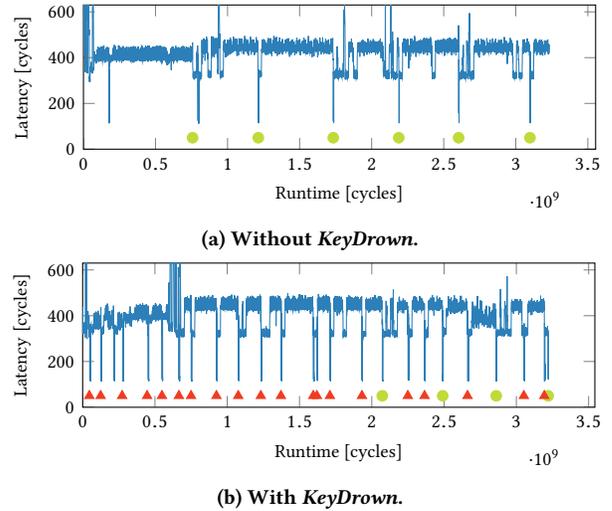

\paragraph{\FlushReload.}
\FlushReload allows an attacker to monitor accesses to memory addresses of a shared library with a very high accuracy.
Figure~\ref{fig:layer2_fr_gdk} shows the result of such an attack against the \texttt{gdk\_keymap\_get\_modifier\_mask} function at address \texttt{0x381c0} of \emph{libgdk-3.so} (v3.20.4 on Ubuntu Linux), the shared library isolating \GTKplus from the windowing system.
This function is executed on every keystroke to retrieve the hardware modifier mask of the windowing system.

Figure~\ref{fig:layer2_fr_kd_disabled} shows the attack when the countermeasure is inactive. 
The attacker measures cache hits on the monitored address whenever a key is pressed and, thus, can spy on the keystroke timings very accurately.
If \KeyDrown is active, as illustrated in Figure~\ref{fig:layer2_fr_kd}, the attacker measures additional cache hits on every injected keystroke and cannot distinguish between them.
For other addresses found using cache template attacks, we made the same observation.
Without \KeyDrown, both profiling and exploiting vulnerable addresses is possible.
With \KeyDrown, we still find all addresses that are loaded into the cache upon keystrokes, however, as we cannot distinguish between real and fake keystrokes we cannot exploit this anymore.
Without \KeyDrown, the precision is \SIx{\ValPrecisionFR} and the \FScore is \SIx{\ValFscoreFR}, which is a +\SI{569.3}{\percent} advantage over an always-one oracle.
If \KeyDrown is active, the precision is lowered to \SIx{\ValPrecisionFRKD} and, thus, the resulting \FScore is \SIx{\ValFscoreFRKD}, which is a (negative) advantage of \SI{-40.2}{\percent} over the always-one oracle.

\paragraph{\PrimeProbe.}
If an attacker cannot use \FlushReload, a fallback to \PrimeProbe is possible.
The disadvantage of a \PrimeProbe attack on the last-level cache is the amount of noise that increases the false-positive rate. 
Prior to this work, there was no successful keystroke attack using \PrimeProbe on the last-level cache. 
We perform the \MultiPrimeProbe attack presented in Section~\ref{sec:new_attack_vectors} to attack keystroke timings. 

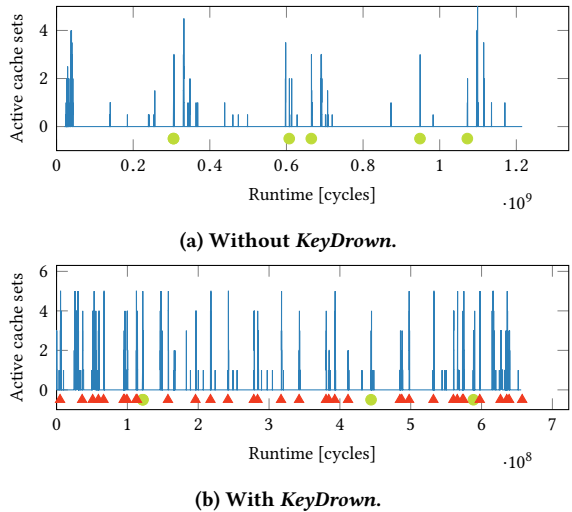
\begin{figure}[t!]
    \centering
    \begin{subfigure}[t]{0.99\hsize}
      \centering
      \begin{tikzpicture}
\begin{axis}[
mlineplot,
style={font=\footnotesize},
xlabel={Runtime [cycles]},
ylabel={Active cache sets},
width=1.0\hsize,
xmin=0,
ymax=5,
ymin=-1,
height=3.5cm
]
\addplot+[blue, no markers] table[x=Time,y=Value,col sep=comma] {data/layer1_prime_probe_without_keydrown.csv};
\addplot+[only marks,mark options={draw=green,fill=green},mark=*] table[x=Time,y=Real,col sep=comma] {data/layer1_prime_probe_without_keydrown.csv};
\addplot+[only marks, mark options={draw=red,fill=red},mark=*] table[x=Time,y=Fake,col sep=comma] {data/layer1_prime_probe_without_keydrown.csv};

\end{axis}
\end{tikzpicture}
      \caption{Without \KeyDrown.}
      \label{fig:layer1_pp_kd_disabled}
    \end{subfigure}%

    \begin{subfigure}[t]{0.99\hsize}
      \centering
      \begin{tikzpicture}
\begin{axis}[
mlineplot,
style={font=\footnotesize},
xlabel={Runtime [cycles]},
ylabel={Active cache sets},
width=1.0\hsize,
xmin=0,
ymax=6.3,
ymin=-1,
height=3.5cm
]
\addplot+[blue, no markers] table[x=Time,y=Value,col sep=comma] {data/layer1_prime_probe_with_keydrown.csv};
\addplot+[only marks,mark options={draw=green,fill=green}, mark=*] table[x=Time,y=Real,col sep=comma] {data/layer1_prime_probe_with_keydrown.csv};
\addplot+[only marks, mark options={draw=red,fill=red},mark=triangle*] table[x=Time,y=Fake,col sep=comma] {data/layer1_prime_probe_with_keydrown.csv};

\end{axis}
\end{tikzpicture}
      \caption{With \KeyDrown.}
      \label{fig:layer1_pp_kd}
    \end{subfigure}%

    \caption{\MultiPrimeProbe attack on the \SIx{\ValPPCacheSets} cache sets from \texttt{0x2514250} to \texttt{0x2514390} of \texttt{i8042\_interrupt}. 
    (a) Noise negatively affects the detection of single keystrokes (\RealMarker). 
    (b) With \KeyDrown, the attacker measures even more cache misses on injected keystrokes (\FakeMarker) as well as on real events (\RealMarker) and cannot distinguish between them.}
    \label{fig:layer1_pp_kernel}
\end{figure}

Figure~\ref{fig:layer1_pp_kernel} shows the results of inferring keystrokes by detecting the keyboard interrupt handler's cache activity using \MultiPrimeProbe.
We monitored \SIx{\ValPPCacheSets} cache sets in parallel for a higher noise robustness. 
Without \KeyDrown, the precision is already at a quite low value of \SIx{\ValPrecisionPP} with a recall of only \SIx{\ValRecallPP}, yielding an \FScore of \SIx{\ValFscorePP}, which is an advantage over an always-one oracle of +\SI{440.0}{\percent}. 
Memory accesses to one of the cache sets by any other application cannot be distinguished from a cache set access by the keyboard interrupt handler, causing a high number of false positives.
If we enable \KeyDrown, the precision drops to \SIx{\ValPrecisionPPKD}, as the attacker additionally measures the noise generated by the injected keystrokes. 
The \FScore is then \SIx{\ValFscorePPKD}, which is a (negative) advantage over an always-one oracle of -\SI{27.7}{\percent}. 

\begin{figure}[t!]
    \centering
    \begin{subfigure}[t]{0.99\hsize}
      \centering
      \begin{tikzpicture}
\begin{axis}[
mlineplot,
style={font=\footnotesize},
xlabel={Runtime [cycles]},
ylabel={Active cache sets},
width=1.0\hsize,
xmin=200000000,
xmax=1700000000,
ymax=5,
ymin=-1,
height=3.5cm
]
\addplot+[blue, no markers] table[x=Time,y=Value,col sep=comma] {data/layer3_prime_probe_without_keydrown.csv};
\addplot+[only marks,mark options={draw=green,fill=green},mark=*] table[x=Time,y=Real,col sep=comma] {data/layer3_prime_probe_without_keydrown.csv};
\addplot+[only marks, mark options={draw=red,fill=red},mark=*] table[x=Time,y=Fake,col sep=comma] {data/layer3_prime_probe_without_keydrown.csv};

\end{axis}
\end{tikzpicture}
      \caption{Without \KeyDrown.}
      \label{fig:layer3_pp_kd_disabled}
    \end{subfigure}%

    \begin{subfigure}[t]{0.99\hsize}
      \centering
      \begin{tikzpicture}
\begin{axis}[
mlineplot,
style={font=\footnotesize},
xlabel={Runtime [cycles]},
ylabel={Active cache sets},
width=1.0\hsize,
xmin=   900000000,
xmax=1800000000,
ymax=5,
ymin=-1, 
height=3.5cm
]
\addplot+[blue, no markers] table[x=Time,y=Value,col sep=comma] {data/layer3_prime_probe_with_keydrown.csv};
\addplot+[only marks, mark options={draw=red,fill=red},mark=triangle*] table[x=Time,y=Fake,col sep=comma] {data/layer3_prime_probe_with_keydrown.csv};
\addplot+[only marks,mark options={draw=green,fill=green},mark=*] table[x=Time,y=Real,col sep=comma] {data/layer3_prime_probe_with_keydrown.csv};

\end{axis}
\end{tikzpicture}
      \caption{With \KeyDrown.}
      \label{fig:layer3_pp_kd}
    \end{subfigure}%
    
    \caption{\MultiPrimeProbe attack on the \SIx{\ValPPCacheSets} cache sets corresponding to a password field's buffer within a demo application.
    (a) Noise negatively affects the detection of single keystrokes (\RealMarker). 
    (b) With \KeyDrown, the attacker measures even more cache misses on injected keystrokes (\FakeMarker) as well as on real events (\RealMarker) and cannot distinguish between them.}
    \label{fig:layer3_pp}
\end{figure}
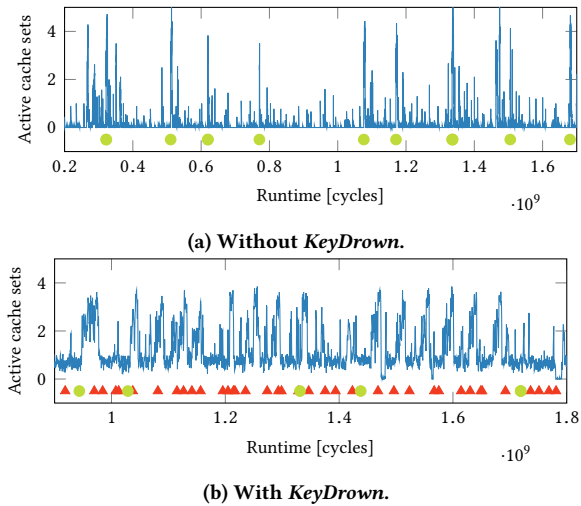

Figure~\ref{fig:layer3_pp} shows the results of mounting a \MultiPrimeProbe attack on the buffer of a password field within a \GTKplus application.
Although there is more noise visible in the traces, we achieve the same precision and \FScore as for the attack on the kernel module when \KeyDrown is disabled. 
If we enable \KeyDrown, the precision drops to \SIx{\ValPrecisionPPKDUser}, which is a bit lower than the precision on the kernel, resulting in an \FScore of \SIx{\ValFscorePPKDUser}, which is again no advantage over an always-one oracle.

\paragraph{Interrupts.}
\KeyDrown also protects against interrupt-based attacks, including our new timing-based attack.
For the attacks based on the \texttt{procfs} interface~\cite{Jana2012,Diao2016}, we measure an average reading interval of \SIx{980} cycles.
With our new attack based on \texttt{rdtsc}, we are able to measure every \SIx{95} cycles  on average, resulting in a probing frequency that is one order of magnitude higher. 

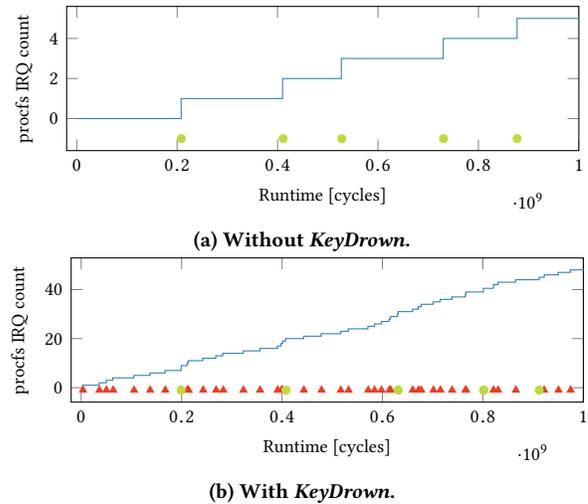
\begin{figure}[t]
\begin{subfigure}[t]{0.99\hsize}
 \centering
 \begin{tikzpicture}
\begin{axis}[
mlineplot,
style={font=\footnotesize},
xlabel={Runtime [cycles]},
ylabel=procfs IRQ count,
width=1.0\hsize,
xmin=-20000000,
xmax=1000000000,
height=3.5cm
]
\addplot+[const plot,blue, no markers] table[x=Time,y=Value,col sep=comma] {data/layer1_procfs_without_keydrown.csv};
\addplot+[only marks, mark options={draw=red,fill=red},mark=*,mark size=1.5] table[x=Time,y=Fake,col sep=comma] {data/layer1_procfs_without_keydrown.csv};
\addplot+[only marks,mark options={draw=green,fill=green},mark=*,mark size=1.5] table[x=Time,y=Real,col sep=comma] {data/layer1_procfs_without_keydrown.csv};

\end{axis}
\end{tikzpicture}
 \caption{Without \KeyDrown.}
 \label{fig:layer1_procfs_wo_kd}
\end{subfigure}%

\begin{subfigure}[t]{0.99\hsize}
 \centering
 \begin{tikzpicture}
\begin{axis}[
mlineplot,
style={font=\footnotesize},
xlabel={Runtime [cycles]},
ylabel=procfs IRQ count,
width=1.0\hsize,
xmin=-20000000,
xmax=1000000000,
height=3.5cm
]
\addplot+[const plot,blue, no markers] table[x=Time,y=Value,col sep=comma] {data/layer1_procfs_with_keydrown.csv};
\addplot+[only marks, mark options={draw=red,fill=red},mark=triangle*,mark size=1.5] table[x=Time,y=Fake,col sep=comma] {data/layer1_procfs_with_keydrown.csv};
\addplot+[only marks,mark options={draw=green,fill=green},mark=*,mark size=1.5] table[x=Time,y=Real,col sep=comma] {data/layer1_procfs_with_keydrown.csv};

\end{axis}
\end{tikzpicture}
 \caption{With \KeyDrown.}
 \label{fig:layer1_procfs_w_kd}
\end{subfigure}%

\caption{\texttt{procfs}-based attack. (a) The attack allows to clearly detect every single
  keyboard interrupt (\RealMarker). (b) With \KeyDrown, the attacker
  measures fake interrupts (\FakeMarker) as well as real interrupts (\RealMarker) and cannot distinguish
between them.}
\label{fig:procfs_attack}
\end{figure}

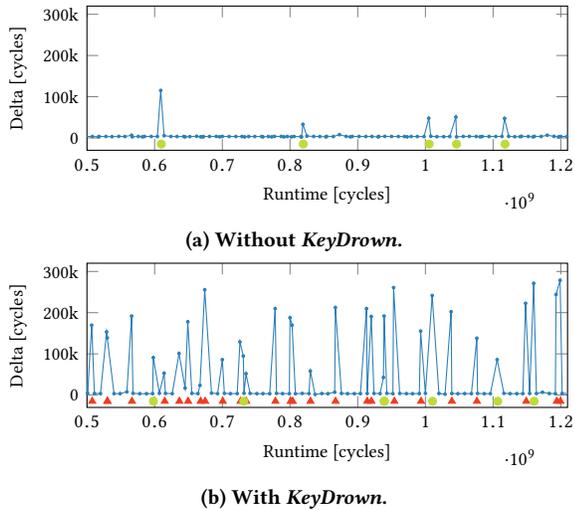
\begin{figure}[t]
\begin{subfigure}[t]{0.99\hsize}
 \centering
 \begin{tikzpicture}
\begin{axis}[
mlineplot,
style={font=\footnotesize},
xlabel={Runtime [cycles]},
ylabel={Delta [cycles]},
width=0.95\hsize,
xmin=500000000,
xmax=1210000000,
scaled y ticks=false,
ytick={0,100000,200000,300000},
yticklabels={0,100k,200k,300k},
ymax=320000,
ymin=-30000,
height=3.5cm
]
\addplot+[blue,thin,mark options={draw=blue,fill=blue},mark=*,mark size=0.5] table[x=Time,y=Value,col sep=comma] {data/layer1_rdtsc_without_keydrown.csv};
\addplot+[only marks, mark options={draw=red,fill=red},mark=triangle*,mark size=1.5] table[x=Time,y=Fake,col sep=comma] {data/layer1_rdtsc_without_keydrown.csv};
\addplot+[only marks,mark options={draw=green,fill=green},mark=*,mark size=1.5] table[x=Time,y=Real,col sep=comma] {data/layer1_rdtsc_without_keydrown.csv};

\end{axis}
\end{tikzpicture}
 \caption{Without \KeyDrown.}
 \label{fig:layer1_rdtsc_wo_kd}
\end{subfigure}%

\begin{subfigure}[t]{0.99\hsize}
 \centering
 \begin{tikzpicture}
\begin{axis}[
mlineplot,
style={font=\footnotesize},
xlabel={Runtime [cycles]},
ylabel={Delta [cycles]},
width=0.95\hsize,
xmin=500000000,
xmax=1210000000,
scaled y ticks=false,
ymax=320000,
ymin=-30000,
ytick={0,100000,200000,300000},
yticklabels={0,100k,200k,300k},
height=3.5cm
]
\addplot+[blue,thin,mark options={draw=blue,fill=blue},mark=*,mark size=0.5] table[x=Time,y=Value,col sep=comma] {data/layer1_rdtsc_with_keydrown.csv};
\addplot+[only marks, mark options={draw=red,fill=red},mark=triangle*,mark size=1.5] table[x=Time,y=Fake,col sep=comma] {data/layer1_rdtsc_with_keydrown.csv};
\addplot+[only marks,mark options={draw=green,fill=green},mark=*,mark size=1.5] table[x=Time,y=Real,col sep=comma] {data/layer1_rdtsc_with_keydrown.csv};


\end{axis}
\end{tikzpicture}
 \caption{With \KeyDrown.}
 \label{fig:layer1_rdtsc_w_kd}
\end{subfigure}%

\caption{\texttt{rdtsc}-based attack. (a) The attack allows to clearly detect every single
  keyboard interrupt (\RealMarker). (b) With \KeyDrown, the attacker
  measures fake interrupts (\FakeMarker) as well as real interrupts (\RealMarker) and cannot distinguish
between them.}
\label{fig:rdtsc_attack}
\end{figure}

Figure~\ref{fig:procfs_attack} and Figure~\ref{fig:rdtsc_attack} illustrate the effect of our countermeasure on the \texttt{procfs}-based interrupt attack and the \texttt{rdtsc}-based attack, respectively. 
Without \KeyDrown, we achieve a precision of \SIx{\ValPrecisionProc} for the \texttt{procfs}-based attack and a precision of \SIx{\ValPrecisionRDTSC} for the \texttt{rdtsc}-based attack, resulting in an \FScore of \SIx{\ValFscoreProc} and \SIx{\ValFscoreRDTSC} respectively. 
Enabling \KeyDrown reduces the precision to \SIx{\ValPrecisionProcKD} and \SIx{\ValPrecisionRDTSCKD} respectively.
Thus, the resulting \FScore is \SIx{\ValFscoreProcKD}, which is exactly the same as the always-one oracle, for the \texttt{procfs}-based attack, and \SIx{\ValFscoreRDTSCKD} for the \texttt{rdtsc}-based attack, which is a (negative) advantage over an always-one oracle of -\SI{3.8}{\percent}.

	\subsection{Requirement \ReqTwo}

\KeyDrown reduced the \FScore of all state-of-the-art attacks such that using the side channel gives an advantage over an always-one oracle of $\leq$\SI{0.0}{\percent}.
An attacker might still be able to combine multiple traces from the same user and build a binary classifier, if the user predictably and repeatedly types the same character sequence. 
Such a classifier may achieve a higher precision and a higher \FScore, as long as there is actually meaningful information in the corresponding traces. 
However, there is a practical limit on the number of traces an attacker can gather from the user, which \ReqTwo estimates to be \SIx{\ValMinTraces} traces.

In our attack scenario, we model a powerful attacker who can take advantage of the following properties: 
\begin{compactenum}
 \item \textbf{Noise-free side channel:} The used side channel is noise-free, \ie only real and fake keystrokes are recorded, no other system noise.
 \item \textbf{Perfect \mbox{(re-)alignment}:} The attacker can detect when a password input starts with a variance as low as the variance of a single inter-keystroke interval.  
 Additionally, the attacker has an alignment-oracle providing perfect re-alignment for the traces after each guessed keystroke. This leads to the same variance for every key instead of an accumulated variance.
 \item \textbf{Known length:} The attacker knows the exact length of the password and expects exactly as many keystrokes. 
\end{compactenum}
This attacker is far stronger than any practical attacker.

We generate simulated traces that fulfill the properties above and calculate the average of the perfectly \mbox{(re-)aligned} traces. 
As our attacker knows the length $n$ of the password, he finds the $n$ most likely positions where a Gaussian distribution with the known inter-keystroke interval variance matches. 
If the expected value $\mu$ of each Gaussian curve is within the variance of the real keystroke, we assume that the number of traces was sufficient to extract the positions of the real keystrokes.

We set the simulated typing variance to $\pm$\SI{40}{\milli\second} which is a bit less than the value reported by Lee~\etal\cite{Lee2015} for trained text sequences. 
In total, we generated \SIx{300000} simulated traces, each containing \SIx{8} keystrokes within \SI{2}{\second}. 
From this set of simulated traces, we evaluated how many randomly chosen traces we have to combine to extract the correct positions of the keystrokes. 
We found that an attacker requires an average of \SIx{\ValReqTraces} traces to extract the correct positions.
This is significantly more than the \SIx{\ValMinTraces} traces deemed to be secure in requirement \ReqTwo.

	\subsection{Requirement \ReqThree}
	
As \KeyDrown fulfills \ReqOne and \ReqTwo, we can be assured that the underlying technique is a working countermeasure. 
However, as the implementation of a countermeasure itself can leak information, we need to make sure that \KeyDrown does not create a new (microarchitectural) side channel in order to satisfy \ReqThree.

\paragraph{First Layer.}
The first layer runs in the kernel and can thus only be attacked using \PrimeProbe. 
Figure~\ref{fig:kernel-module} shows that, in general, we have the same execution flow and data accesses. 
For the few deviations, we prevent any potential cache leakage from non-executed code paths by performing the same memory accesses as if they were executed.
As an attacker cannot distinguish if a cache activity is caused by an execution or a memory read, the module's cache activity does not leak additional information to an attacker. 
We investigated the cache activity on the cache sets used by the \KeyDrown kernel module in a \PrimeProbe attack and found no leakage from our module.

\paragraph{Second Layer.}
To make use of the same noise as in the first layer, the second layer listens to the \texttt{/dev/input/event0} pseudo-file containing all keyboard events.
This file is not world-readable but only readable by users part of the \texttt{input} user group. 
Thus, this layer runs under a separate \textit{keydrown} user with default limited privileges and additional access to this file. 

As the second layer is a user space binary, an attacker could theoretically mount a \FlushReload attack against the second layer. 
However, attacking the second layer does not result in any additional information. 
The second layer does not know whether an event is generated from a real or an injected keystroke. 
For every event, a random printable character is sent to the hidden window. 
Thus, the execution path for printable characters is always active and the attacker cannot learn any additional information from attacking the second layer. 
The same is also true for \PrimeProbe, even a successful attack does not provide additional information. 
We investigated the cache activity of the \KeyDrown shared library parts and the \KeyDrown user space binary using a template attack and did not find any leakage.

\paragraph{Third Layer.}

The third layer builds upon the second layer, and thus the same argumentation as for the second layer holds. 
An attacker cannot distinguish real and injected keystrokes in the second layer as all events are merged within the kernel. 
As the third layer relies on the same source as the second layer, there is also no leakage from the third layer. 
Thus, any attack on the third layer does not give an attacker any advantage over any other attack.
We investigated the cache activity of the control flow and data accesses up to the point where the input is stored in the buffer in a \PrimeProbe attack and found no leakage.

	\subsection{Performance}

On the x86 architecture, we evaluate the performance impacts of running our \KeyDrown implementation on standard Ubuntu 16.10. 
We use \textit{lmbench}~\cite{McVoy1996}, a set of micro benchmarks for performance analysis of UNIX systems, and \textit{PARSEC 3.0}~\cite{Bienia2008}, a benchmark suite intended to simulate a realistic workload on multicore systems. 

The lmbench results for the latency benchmarks show a performance overhead of \SI{\ValLmbenchLatency}{\percent}. 
However, as the execution time of the lmbench benchmarks is in the range of microseconds to nanoseconds, the overhead does not allow for definite conclusions about the overall system performance. 
Still, we can see that the injected interrupts have only a small impact on the kernel performance. 

To measure the overall performance, we run the PARSEC 3.0 benchmark with different numbers of cores. 
The average performance overhead over all measurements for any number of cores is \SI{\ValPARSEC}{\percent}. 
For workloads that do not use all cores, the performance impact is only \SI{\ValPARSECOne}{\percent} for one core and \SI{\ValPARSECTwo}{\percent} for two cores. 
Only if the CPU is under heavy load, we observe a higher performance impact of \SI{\ValPARSECFour}{\percent} when running the benchmarks on all cores. 

On ARM, we evaluate the battery consumption of \KeyDrown. 
We measure the power consumption in three different scenarios, always over the timespan of \SI{5}{\minute}. 
First, if the screen is off, our fake interrupts are completely disabled, and thus, \KeyDrown does not increase the power consumption if the mobile phone is not used. 
Second, if the screen is turned on, but the keyboard is not shown. 
In this case, \KeyDrown increases the power consumption slightly by \SI{\ValBatteryIdle}{\percent}. 
Third, if the keyboard is shown, the power consumption with \KeyDrown increases by \SI{\ValBatteryKeyboard}{\percent}. 
However, as most of the time the keyboard is not shown, \KeyDrown does not have great impacts on the overall power consumption. 

Note that all the performance measurements were done using the proof-of-concept. 
We expect that the proof-of-concept can be considerably improved in terms of performance overhead and battery usage by not injecting the fake interrupts all the time but only while the user is actually entering text.

\subsection{Other Attacks}
While we already demonstrated that the most powerful side-channel attacks are mitigated, we discuss three other attacks subsequently.
The \PrimeProbe side channel results from the victim program evicting a cache line of the attacker.
As the last-level cache is inclusive, any eviction from the last-level cache also evicts this line from the L1 cache.
However, if a cache line is evicted from the L1 cache it may still be in the last-level cache.
In this case the attacker would miss the eviction and thus the targeted event.
In our evaluation we find that the recall is very close to \num{1} in all cases.
This means that we do not miss any events.
Hence, there is no additional information that an attacker could gain from a \PrimeProbe attack on the L1 cache.
Consequently, evaluating \PrimeProbe on the last-level cache is sufficient to conclude that \PrimeProbe on the L1 cache does not leak additional information.

The DRAMA side-channel attack presented by Pessl~\etal\cite{Pessl2016} results from a massive number of secret-dependent memory accesses that lead to heavy cache thrashing, \ie the victim program accesses lots of memory locations that are mapped to the same cache lines. 
It is therefore unclear whether or not \KeyDrown protects against DRAMA. In particular, it does not protect against the specific attack against keystrokes in the Firefox address bar (\cf Section~\ref{sec:limitations}). 

Vila~\etal\cite{Vila2017} showed a keystroke timing attack based on the event queue of the Chrome browser. 
They state that the leakage is due to the time it takes Chrome to enqueue and dispatch every keystroke event. 
Thus, this attack is also out-of-scope (\cf Section~\ref{sec:limitations}) for \KeyDrown. 
Surprisingly, we noticed that \KeyDrown adds measurable noise to their attack, which makes it difficult to see the real keystroke timings. 
Thus, we concluded that the authors do not only detect timing differences in the event queue, but also see hardware interrupts. 
To confirm our hypothesis, we tested their attack on Firefox, where it---to a lesser extent---also works, and where \KeyDrown was able to fully prevent it. 

\section{Limitations and Future Work}\label{sec:limitations}
\KeyDrown mitigates interrupt-based attacks as well as microarchitectural attacks on keystrokes and keystroke timings in general. 
This includes even the application layer without changing an existing application if either:
\begin{itemize}[noitemsep,topsep=0pt,parsep=0pt,partopsep=0pt,leftmargin=16pt]
 \item the input is processed only after the user finished entering the text, \eg by pressing a button on a login form, and there is no immediate action when a key is pressed, \eg as it is the case in password fields or simple text input fields,
 \item the application is designed to remove side-channel information.
\end{itemize}
However, \KeyDrown \emph{does not} prevent all possible side-channel attacks on keyboard input. 
Depending on the implementation of the application, the application layer might still leak timing information. 
Examples include but are not limited to:
\begin{itemize}[noitemsep,topsep=0pt,parsep=0pt,partopsep=0pt,leftmargin=16pt]
 \item key press/release handlers reacting on every keystroke and executing code which might be detected due to CPU utilization~\cite{Jana2012}, network traffic~\cite{Zhang2009}, or screen redraws~\cite{Diao2016},
 \item operations that are executed after every text update, such as \eg autocomplete or live search features~\cite{Pessl2016}.
\end{itemize}

Other side-channel information may allow inferring keyboard input directly. 
For instance, various sensors like the accelerometer~\cite{DBLP:conf/uss/CaiC11} have been successfully exploited to infer keyboard input on mobile devices. 
While such specific attacks can be thwarted by restricting access to specific resources and by injecting noise in the sensor values~\cite{DBLP:conf/wisec/ShresthaMS16}, we consider these attacks out of scope in this paper. 

We demonstrate \KeyDrown protection against interrupt-based attacks and microarchitectural attacks on keystrokes as well as touch events. 
However, swipe movements are not protected as their interrupt rate is too high.
While this is not a problem in the case of a password input---if a password can be swiped and thus pasted from a dictionary, there is little to protect---it is future work to investigate how to extend \KeyDrown to protect swipe movements in general.

Furthermore, our novel side channels emphasize the necessity to deploy \KeyDrown widely. 
\MultiPrimeProbe attacks provide a significantly higher accuracy than previous \PrimeProbe attacks on dynamic memory and kernel memory. 
It is likely that \MultiPrimeProbe works similarly in cloud systems and thus allows highly accurate attacks like keystroke timing attacks across virtual machine boundaries.

\section{Conclusion}\label{sec:conclusion}
Keystrokes are processed on many different layers of the software stack and are thus not entirely covered by previously proposed defense mechanisms.
In this article we presented \KeyDrown, a novel defense mechanism that mitigates keystroke timing attacks.
\KeyDrown injects a larger number of fake keystrokes on the kernel level and propagates them---through all layers of the software stack---up to the user space application. 
A careful design and implementation of this countermeasure ensures that all software routines involved in the processing of a keystroke are loaded, 
irrespective of whether a real or a fake keystroke is processed. 
Thereby, \KeyDrown mitigates interrupt-based attacks, \PrimeProbe attacks, and \FlushReload attacks on the entire software stack. 
With \KeyDrown, an attacker cannot distinguish fake from real keystrokes in practice anymore.
Our evaluation shows that \KeyDrown eliminates any advantage an attacker can gain from side channels, \ie $\leq$\SI{0.0}{\percent} advantage over an always-one oracle, and thus successfully mitigates keystroke timing attacks.

\bibliographystyle{ACM-Reference-Format}
\bibliography{main}

\FloatBarrier

\appendix

\begin{figure}[t]
    \centering
    \begin{subfigure}[t]{0.99\hsize}
      \centering
      \begin{tikzpicture}
\begin{axis}[
mlineplot,
style={font=\footnotesize},
xlabel={Runtime [ns]},
ylabel={Latency [cycles]},
width=0.95\hsize,
xmin=0,
ymax=1200,
xmax=3026000000,
ymin=0,
height=3.5cm
]
\addplot+[blue, no markers,restrict x to domain=0:3026000000] table[x=Time,y=Value,col sep=comma] {data/layer2_flush_reload_without_keydrown_nexus.csv};

\addplot+[only marks, mark options={draw=red,fill=red},mark=triangle*,restrict x
  to domain=0:3026000000] table[x=Time,y=Fake,col sep=comma]{data/layer2_flush_reload_without_keydrown_nexus.csv};

\addplot+[only marks,mark options={draw=green,fill=green}, mark=*,restrict x to
  domain=0:3026000000] table[x=Time,y=Real,col sep=comma]{data/layer2_flush_reload_without_keydrown_nexus.csv};

\end{axis}
\end{tikzpicture}
      \caption{Without \KeyDrown.}
      \label{fig:layer2_fr_kd_disabled_nexus}
    \end{subfigure}%

    \begin{subfigure}[t]{0.99\hsize}
      \centering
      \begin{tikzpicture}
\begin{axis}[
mlineplot,
style={font=\footnotesize},
xlabel={Runtime [ns]},
ylabel={Latency [cycles]},
width=0.95\hsize,
xmin=0,
ymax=1200,
xmax=3545000000,
ymin=0,
height=3.5cm
]
\addplot+[blue, no markers,restrict x to domain=0:3545000000] table[x=Time,y=Value,col sep=comma] {data/layer2_flush_reload_with_keydrown_nexus.csv};

\addplot+[only marks, mark options={draw=red,fill=red},mark=triangle*,restrict x
  to domain=0:3545000000] table[x=Time,y=Fake,col sep=comma]{data/layer2_flush_reload_with_keydrown_nexus.csv};

\addplot+[only marks,mark options={draw=green,fill=green}, mark=*,restrict x to
  domain=0:3545000000] table[x=Time,y=Real,col sep=comma]{data/layer2_flush_reload_with_keydrown_nexus.csv};

\end{axis}
\end{tikzpicture}
      \caption{With \KeyDrown.}
      \label{fig:layer2_fr_kd_nexus}
    \end{subfigure}%

    \caption{\FlushReload attack on address \texttt{0xfb5a} of
      \texttt{libinput.so} on the \Nexus. (a) The attack allows to clearly detect every single
      keystroke (\RealMarker). (b) With \KeyDrown, the attacker
      measures cache hits on injected keystrokes (\FakeMarker) as well as on real events (\RealMarker) and cannot distinguish
    between them.}
    \label{fig:layer2_fr_libinput_nexus}
\end{figure}
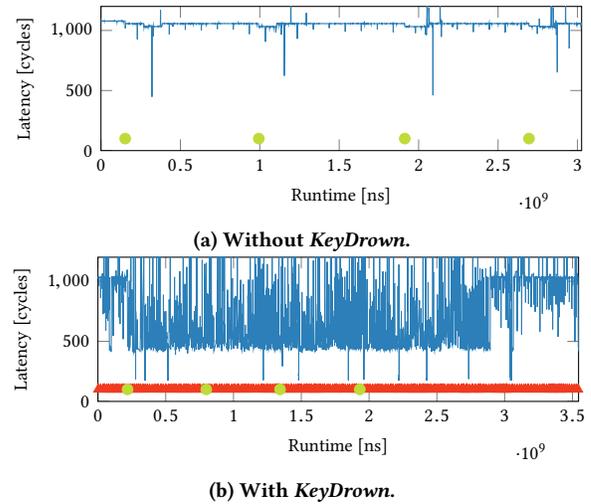

\begin{figure}[t]
\vspace{1em}
\begin{subfigure}[t]{0.99\hsize}
 \centering
 \begin{tikzpicture}
\begin{axis}[
mlineplot,
style={font=\footnotesize},
xlabel={Runtime [ns]},
ylabel=procfs IRQ count,
width=1.0\hsize,
xmin=500000000,
xmax=4000000000,
height=3.5cm
]
\addplot+[const plot,blue, no markers] table[x=Time,y=Value,col sep=comma] {data/layer1_procfs_without_keydrown_nexus.csv};
\addplot+[only marks,mark options={draw=green,fill=green},mark=*,mark size=1.5] table[x=Time,y=Real,col sep=comma] {data/layer1_procfs_without_keydrown_nexus.csv};

\end{axis}
\end{tikzpicture}
 \caption{Without \KeyDrown.}
 \label{fig:layer1_procfs_wo_kd_nexus}
\end{subfigure}
\begin{subfigure}[t]{0.99\hsize}
 \centering
 \begin{tikzpicture}
\begin{axis}[
mlineplot,
style={font=\footnotesize},
xlabel={Runtime [ns]},
ylabel=procfs IRQ count,
width=1.0\hsize,
xmin=100000000,
xmax=800000000,
height=3.5cm
]
\addplot+[const plot,blue, no markers] table[x=Time,y=Value,col sep=comma] {data/layer1_procfs_with_keydrown_nexus.csv};
\addplot+[only marks, mark options={draw=red,fill=red},mark=triangle*,mark size=1.5] table[x=Time,y=Fake,col sep=comma] {data/layer1_procfs_with_keydrown_nexus.csv};
\addplot+[only marks,mark options={draw=green,fill=green},mark=*,mark size=1.5] table[x=Time,y=Real,col sep=comma] {data/layer1_procfs_with_keydrown_nexus.csv};

\end{axis}
\end{tikzpicture}
 \caption{With \KeyDrown.}
 \label{fig:layer1_procfs_w_kd_nexus}
\end{subfigure}
\caption{\texttt{procfs}-based attack on the \Nexus. (a) The attack allows to clearly detect every single
  touchscreen interrupt (\RealMarker). (b) With \KeyDrown, the attacker measures fake interrupts (\FakeMarker) as well as real interrupts (\RealMarker) and cannot distinguish
between them.}
\label{fig:procfs_attack_nexus}
\end{figure}
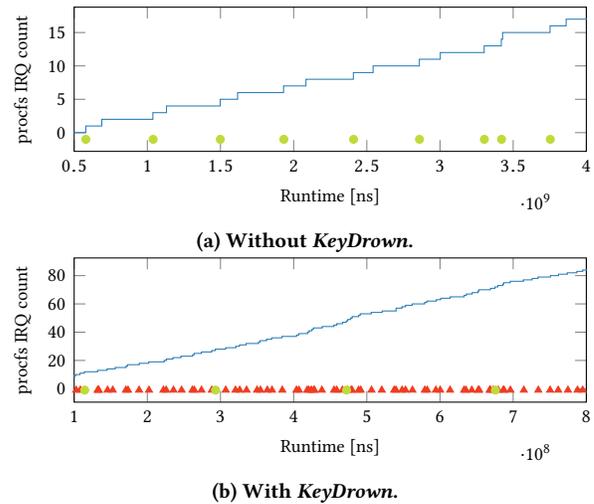

\section{Effect of \KeyDrown on Different Architectures}\label{app:comparison}
In this section, we compare the accuracy of four different side channels with and without \KeyDrown: \texttt{procfs}, \texttt{rdtsc}, \FlushReload, and \PrimeProbe on the last-level cache.
We compare these attacks on three different architectures: a Lenovo ThinkPad T460s with an Intel Core i5-6200U, an LG \Nexus (ARMv7), and a \OnePlus (ARMv8).
Table~\ref{tab:fscore_large} summarizes the \FScore{s} for all attacks with and without \KeyDrown.
\KeyDrown prevents keystroke timing attacks in all cases when considering single-trace attacks only. 

\setlength{\aboverulesep}{0pt}
\setlength{\belowrulesep}{0pt}
\begin{table*}
	\centering
\caption{\FScore without and with \KeyDrown for state-of-the-art attacks.}\label{tab:fscore_large}
\rowcolors{2}{white}{lightgray!40}
\begin{tabular}{p{3.2cm}p{3.8cm}cc}
\toprule
Device &	Side Channel & without \KeyDrown & with \KeyDrown \\ 
\midrule	
ThinkPad T460s & \texttt{procfs} & \SIx{\ValFscoreProc} & \SIx{\ValFscoreProcKD} \\ 
LG \Nexus & \texttt{procfs} & \SIx{\ValFscoreProcNexus} & \SIx{\ValFscoreProcKDNexus} \\ 
\OnePlus & \texttt{procfs} & \SIx{\ValFscoreProcOP} & \SIx{\ValFscoreProcKDOP} \\ 
\midrule
ThinkPad T460s & Interrupt-timing (\texttt{rdtsc}) & \SIx{\ValFscoreRDTSC} & \SIx{\ValFscoreRDTSCKD} \\ 
LG \Nexus & Interrupt-timing & \SIx{\ValFscoreRDTSCNexus} & \SIx{\ValFscoreRDTSCKDNexus} \\ 
\OnePlus & Interrupt-timing & \SIx{\ValFscoreRDTSCOP} & \SIx{\ValFscoreRDTSCKDOP} \\ 
\midrule
ThinkPad T460s & \FlushReload	& \SIx{\ValFscoreFR} & \SIx{\ValFscoreFRKD} \\ 
LG \Nexus & \FlushReload	& \SIx{\ValFscoreFRNexus} & \SIx{\ValFscoreFRKDNexus} \\ 
\OnePlus & \FlushReload	& \SIx{\ValFscoreFROP} & \SIx{\ValFscoreFRKDOP} \\ 
\midrule
ThinkPad T460s & \PrimeProbe on LLC	& \SIx{\ValFscorePP} & \SIx{\ValFscorePPKD} \\ 
LG \Nexus & \PrimeProbe on LLC	& \SIx{\ValFscorePPNexus} & \SIx{\ValFscorePPKDNexus} \\ 
\OnePlus & \PrimeProbe on LLC	& \SIx{\ValFscorePPOP} & \SIx{\ValFscorePPKDOP} \\ 
\bottomrule
\end{tabular} 
\end{table*}

\section{\Nexus}\label{app:nexus}

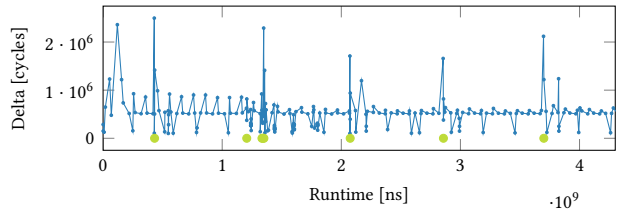
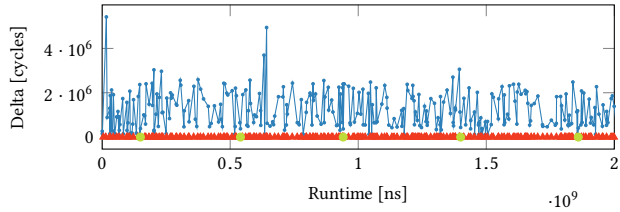
\begin{figure}[t]
\begin{subfigure}[t]{0.99\hsize}
 \centering
 \begin{tikzpicture}
\begin{axis}[
mlineplot,
style={font=\footnotesize},
xlabel={Runtime [ns]},
ylabel={Delta [cycles]},
width=1.0\hsize,
xmin=000000000,
xmax=4300000000,
scaled y ticks=false,
height=3.5cm
]

\addplot+[blue,thin,mark options={draw=blue,fill=blue},mark=*,mark size=0.5] table[x=Time,y=Value,col sep=comma] {data/layer1_rdtsc_without_keydrown_nexus.csv};
\addplot+[only marks, mark options={draw=red,fill=red},mark=triangle*,mark size=1.5] table[x=Time,y=Fake,col sep=comma] {data/layer1_rdtsc_without_keydrown_nexus.csv};
\addplot+[only marks,mark options={draw=green,fill=green},mark=*,mark size=1.5] table[x=Time,y=Real,col sep=comma] {data/layer1_rdtsc_without_keydrown_nexus.csv};

\end{axis}
\end{tikzpicture}
 \caption{Without \KeyDrown.}
 \label{fig:layer1_rdtsc_wo_kd_nexus}
\end{subfigure}
\begin{subfigure}[t]{0.99\hsize}
 \centering
 \begin{tikzpicture}
\begin{axis}[
mlineplot,
style={font=\footnotesize},
xlabel={Runtime [ns]},
ylabel={Delta [cycles]},
width=1.0\hsize,
xmin=0,
xmax=2000000000,
scaled y ticks=false,
height=3.5cm
]
\addplot+[blue,thin,mark options={draw=blue,fill=blue},mark=*,mark size=0.5] table[x=Time,y=Value,col sep=comma] {data/layer1_rdtsc_with_keydrown_nexus.csv};
\addplot+[only marks, mark options={draw=red,fill=red},mark=triangle*,mark size=1.5] table[x=Time,y=Fake,col sep=comma] {data/layer1_rdtsc_with_keydrown_nexus.csv};
\addplot+[only marks,mark options={draw=green,fill=green},mark=*,mark size=1.5] table[x=Time,y=Real,col sep=comma] {data/layer1_rdtsc_with_keydrown_nexus.csv};


\end{axis}
\end{tikzpicture}
 \caption{With \KeyDrown.}
 \label{fig:layer1_rdtsc_w_kd_nexus}
\end{subfigure}
\caption{Timing-based attack on the \Nexus. (a) The attack allows to clearly detect every single
  touchscreen interrupt (\RealMarker). (b) With \KeyDrown, the attacker measures fake interrupts (\FakeMarker) as well as real interrupts (\RealMarker) and cannot distinguish
between them.}
\label{fig:rdtsc_attack_nexus}
\end{figure}

We performed our experiments on the touchscreen soft-keyboard of the \Nexus.
Figure~\ref{fig:layer2_fr_libinput_nexus} shows a \FlushReload attack on
\texttt{libinput.so}.
Without \KeyDrown, the precision is \SIx{\ValPrecisionFRNexus} and the \FScore is thus \SIx{\ValFscoreFRNexus}.
If \KeyDrown is active, the precision is lowered to \SIx{\ValPrecisionFRKDNexus} and, thus, the resulting \FScore of \SIx{\ValFscoreFRKDNexus} means a $\leq$\SI{-86.5}{\percent} advantage over an always-one oracle.

Figure~\ref{fig:procfs_attack_nexus} and Figure~\ref{fig:rdtsc_attack_nexus} show a \texttt{procfs}-based interrupt attack and a timing-based attack, both on the \Nexus. 
Without \KeyDrown, we achieve a precision of \SIx{\ValPrecisionProcNexus} for the \texttt{procfs}-based attack and a precision of \SIx{\ValPrecisionRDTSCNexus} for the timing-based attack, resulting in an \FScore of \SIx{\ValFscoreProcNexus} and \SIx{\ValFscoreRDTSCNexus} respectively. 
Enabling \KeyDrown reduces the precision to only \SIx{\ValPrecisionProcKD} and \SIx{\ValPrecisionRDTSCKD} respectively.
Thus, the resulting \FScore is \SIx{\ValFscoreProcKDNexus} for the \texttt{procfs}-based attack, and \SIx{\ValFscoreRDTSCKDNexus} for the timing-based attack, which is an advantage of $\leq$\SI{0.0}{\percent} over an always-one oracle.

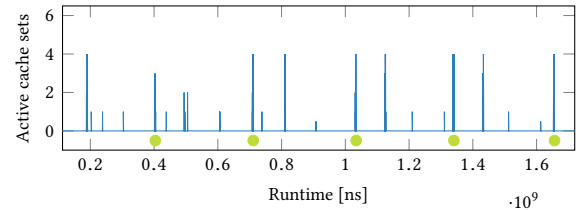
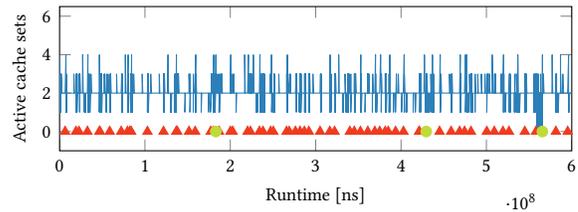
\begin{figure}[!t]
    \centering
    \begin{subfigure}[t]{0.99\hsize}
      \centering
      \begin{tikzpicture}
\begin{axis}[
mlineplot,
style={font=\footnotesize},
xlabel={Runtime [ns]},
ylabel={Active cache sets},
width=1.0\hsize,
xmin=112800000,
xmax=1719611442,
ymax=6.5,
ymin=-1,
height=3.5cm
]
\addplot+[blue, no markers] table[x=Time,y=Value,col sep=comma] {data/layer1_prime_probe_without_keydrown_nexus.csv};
\addplot+[only marks,mark options={draw=green,fill=green}, mark=*] table[x=Time,y=Real,col sep=comma] {data/layer1_prime_probe_without_keydrown_nexus.csv};
\addplot+[only marks, mark options={draw=red,fill=red},mark=triangle*] table[x=Time,y=Fake,col sep=comma] {data/layer1_prime_probe_without_keydrown_nexus.csv};

\end{axis}
\end{tikzpicture}
      \caption{Without \KeyDrown.}
      \label{fig:layer1_pp_kd_disabled_nexus}
    \end{subfigure}%

    \begin{subfigure}[t]{0.99\hsize}
      \centering
      \begin{tikzpicture}
\begin{axis}[
mlineplot,
style={font=\footnotesize},
xlabel={Runtime [ns]},
ylabel={Active cache sets},
width=1.0\hsize,
xmin=0,
xmax=600000000,
ymax=6.5,
ymin=-1,
height=3.5cm
]
\addplot+[blue, no markers] table[x=Time,y=Value,col sep=comma] {data/layer1_prime_probe_with_keydrown_nexus.csv};
\addplot+[only marks, mark options={draw=red,fill=red},mark=triangle*] table[x=Time,y=Fake,col sep=comma] {data/layer1_prime_probe_with_keydrown_nexus.csv};
\addplot+[only marks,mark options={draw=green,fill=green}, mark=*] table[x=Time,y=Real,col sep=comma] {data/layer1_prime_probe_with_keydrown_nexus.csv};

\end{axis}
\end{tikzpicture}
      \caption{With \KeyDrown.}
      \label{fig:layer1_pp_kd_nexus}
    \end{subfigure}%

    \caption{\MultiPrimeProbe attack on the \SIx{\ValPPCacheSets} cache sets from \texttt{0x382659be} to \texttt{0x38265abe} of \texttt{touch\_irq\_handler} on the \Nexus.
    (a) Noise negatively affects the detection of single keystrokes (\RealMarker). 
    (b) With \KeyDrown, the attacker measures even more cache misses on injected keystrokes (\FakeMarker) as well as on real events (\RealMarker) and cannot distinguish between them.}
    \label{fig:layer1_pp_kernel_nexus}
\end{figure}

Figure~\ref{fig:layer1_pp_kernel_nexus} shows the results of inferring keystrokes by detecting the touchscreen interrupt handler's cache activity using \MultiPrimeProbe on the \Nexus.
We monitored \SIx{\ValPPCacheSetsNexus} cache sets in parallel for a higher noise robustness.
Without \KeyDrown, the precision is already at a quite low value of \SIx{\ValPrecisionPPNexus} with a recall of only \SIx{\ValRecallPPNexus}, as an access to one of the cache sets by any other application cannot be distinguished from a cache set access by the touchscreen interrupt handler.
Thus, this attack has a high number of false positives. 
If we enable \KeyDrown, the precision drops to \SIx{\ValPrecisionPPKDNexus}, as the attacker additionally measures the noise generated by the injected keystrokes. 
Thus, the \FScore is \SIx{\ValFscorePPKDNexus}.

\section{\OnePlus}\label{app:op3}

\begin{figure}[t]
    \centering
    \begin{subfigure}[t]{0.99\hsize}
      \centering
      \begin{tikzpicture}
\begin{axis}[
mlineplot,
style={font=\footnotesize},
xlabel={Runtime [ns]},
ylabel={Latency [cycles]},
width=1.0\hsize,
xmin=1000000000,
xmax=7832000000,
ymax=1600,
ymin=0,
height=3.5cm
]
\addplot+[blue, no markers,restrict x to domain=1000000000:4174402814] table[x=Time,y=Value,col sep=comma]{data/layer2_flush_reload_without_keydrown_oneplus3t.csv};
\addplot+[blue, no markers,restrict x to domain=4174402814:7832000000] table[x=Time,y=Value,col sep=comma]{data/layer2_flush_reload_without_keydrown_oneplus3t.csv};
\addplot+[only marks, mark options={draw=red,fill=red},mark=*,restrict x to domain=1000000000:4174402814] table[x=Time,y=Fake,col sep=comma]{data/layer2_flush_reload_without_keydrown_oneplus3t.csv};
\addplot+[only marks, mark options={draw=red,fill=red},mark=*,restrict x to domain=4174402814:7832000000] table[x=Time,y=Fake,col sep=comma]{data/layer2_flush_reload_without_keydrown_oneplus3t.csv};
\addplot+[only marks,mark options={draw=green,fill=green},mark=*,restrict x to domain=1000000000:4174402814] table[x=Time,y=Real,col sep=comma] {data/layer2_flush_reload_without_keydrown_oneplus3t.csv};
\addplot+[only marks,mark options={draw=green,fill=green},mark=*,restrict x to domain=4174402814:7832000000] table[x=Time,y=Real,col sep=comma] {data/layer2_flush_reload_without_keydrown_oneplus3t.csv};

\end{axis}
\end{tikzpicture}
      \caption{Without \KeyDrown.}
      \label{fig:layer2_fr_kd_disabled_oneplus3t}
    \end{subfigure}%

    \begin{subfigure}[t]{0.99\hsize}
      \centering
      \begin{tikzpicture}
\begin{axis}[
mlineplot,
style={font=\footnotesize},
xlabel={Runtime [ns]},
ylabel={Latency [cycles]},
width=1.0\hsize,
xmin=0,
ymax=1600,
xmax=7832000000,
ymin=0,
height=3.5cm
]
\addplot+[blue, no markers,restrict x to domain=0:3916000000] table[x=Time,y=Value,col sep=comma] {data/layer2_flush_reload_with_keydrown_oneplus3t.csv};
\addplot+[blue, no markers,restrict x to domain=3916000001:7832000000] table[x=Time,y=Value,col sep=comma] {data/layer2_flush_reload_with_keydrown_oneplus3t.csv};
\addplot+[only marks, mark options={draw=red,fill=red},mark=triangle*,restrict x to domain=0:3916000000] table[x=Time,y=Fake,col sep=comma]{data/layer2_flush_reload_with_keydrown_oneplus3t.csv};
\addplot+[only marks, mark options={draw=red,fill=red},mark=triangle*,restrict x to domain=3916000001:7832000000] table[x=Time,y=Fake,col sep=comma]{data/layer2_flush_reload_with_keydrown_oneplus3t.csv};
\addplot+[only marks,mark options={draw=green,fill=green}, mark=*,restrict x to domain=0:3916000000] table[x=Time,y=Real,col sep=comma]{data/layer2_flush_reload_with_keydrown_oneplus3t.csv};
\addplot+[only marks,mark options={draw=green,fill=green}, mark=*,restrict x to domain=3916000001:7832000000] table[x=Time,y=Real,col sep=comma]{data/layer2_flush_reload_with_keydrown_oneplus3t.csv};

\end{axis}
\end{tikzpicture}
      \caption{With \KeyDrown.}
      \label{fig:layer2_fr_kd_oneplus3t}
    \end{subfigure}%

    \caption{\FlushReload attack on address \texttt{0x28ec0} of
      \texttt{libflinger.so} on the \OnePlus. (a) The attack allows to clearly detect every single
      keystroke (\RealMarker). (b) With \KeyDrown, the attacker measures cache hits on injected keystrokes (\FakeMarker) as well as on real events (\RealMarker) and cannot distinguish
    between them.}
    \label{fig:layer2_fr_gdk_oneplus3t}
\end{figure}
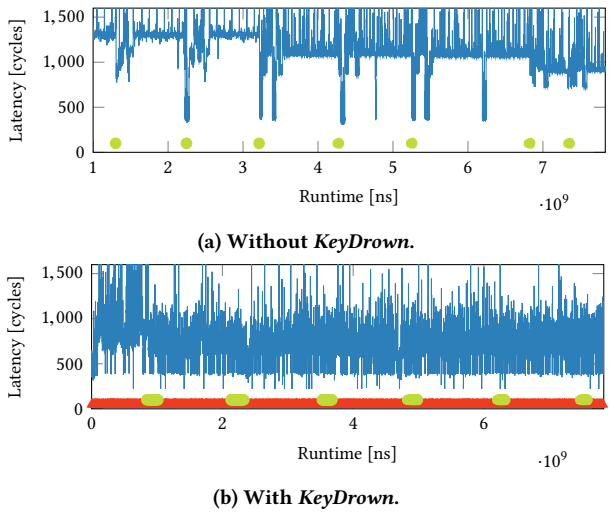

We performed our experiments on the touchscreen soft-keyboard of the \OnePlus.
Figure~\ref{fig:layer2_fr_gdk_oneplus3t} shows a \FlushReload attack on \texttt{libflinger.so} on the \OnePlus.
Without \KeyDrown, the precision is \SIx{\ValPrecisionFROP} and the \FScore is thus \SIx{\ValFscoreFROP}.
If \KeyDrown is active, the precision is lowered to \SIx{\ValPrecisionFRKDOP} and, thus, the resulting \FScore of \SIx{\ValFscoreFRKDOP} means a $\leq$\SI{-32.5}{\percent} advantage over an always-one oracle.

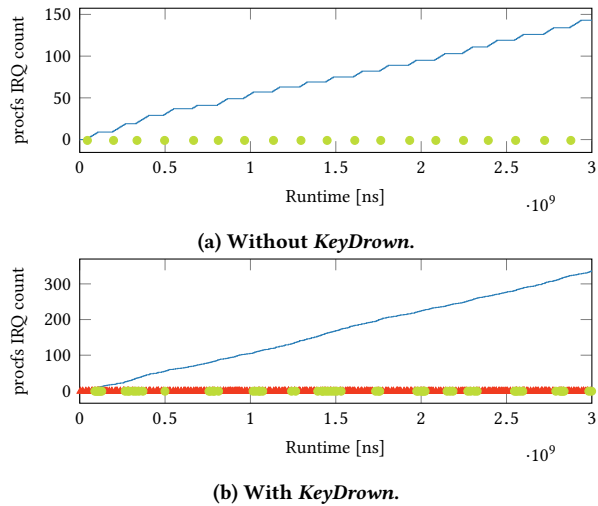
\begin{figure}[t]
\begin{subfigure}[t]{0.99\hsize}
 \centering
 \begin{tikzpicture}
\begin{axis}[
mlineplot,
style={font=\footnotesize},
xlabel={Runtime [ns]},
ylabel=procfs IRQ count,
width=1.0\hsize,
xmin=0,
xmax=3000000000,
height=3.5cm
]
\addplot+[const plot,blue, no markers] table[x=Time,y=Value,col sep=comma] {data/layer1_procfs_without_keydrown_oneplus3t.csv};
\addplot+[only marks,mark options={draw=green,fill=green},mark=*,mark size=1.5] table[x=Time,y=Real,col sep=comma] {data/layer1_procfs_without_keydrown_oneplus3t.csv};

\end{axis}
\end{tikzpicture}
 \caption{Without \KeyDrown.}
 \label{fig:layer1_procfs_wo_kd_oneplus3t}
\end{subfigure}
\begin{subfigure}[t]{0.99\hsize}
 \centering
 \begin{tikzpicture}
\begin{axis}[
mlineplot,
style={font=\footnotesize},
xlabel={Runtime [ns]},
ylabel=procfs IRQ count,
width=1.0\hsize,
xmin=0,
xmax=3000000000,
height=3.5cm
]
\addplot+[const plot,blue, no markers] table[x=Time,y=Value,col sep=comma] {data/layer1_procfs_with_keydrown_oneplus3t.csv};
\addplot+[only marks, mark options={draw=red,fill=red},mark=triangle*,mark size=1.5] table[x=Time,y=Fake,col sep=comma] {data/layer1_procfs_with_keydrown_oneplus3t.csv};
\addplot+[only marks,mark options={draw=green,fill=green},mark=*,mark size=1.5] table[x=Time,y=Real,col sep=comma] {data/layer1_procfs_with_keydrown_oneplus3t.csv};

\end{axis}
\end{tikzpicture}
 \caption{With \KeyDrown.}
 \label{fig:layer1_procfs_w_kd_oneplus3t}
\end{subfigure}
\caption{\texttt{procfs}-based attack on the \OnePlus. (a) Without \KeyDrown, one can clearly see the interrupts occurring on real key events. (b) With \KeyDrown enabled, the interrupts (\RealMarker) are hidden in the noise of injected interrupts (\FakeMarker).}
\vspace{1cm}
\label{fig:procfs_attack_oneplus3t}
\end{figure}

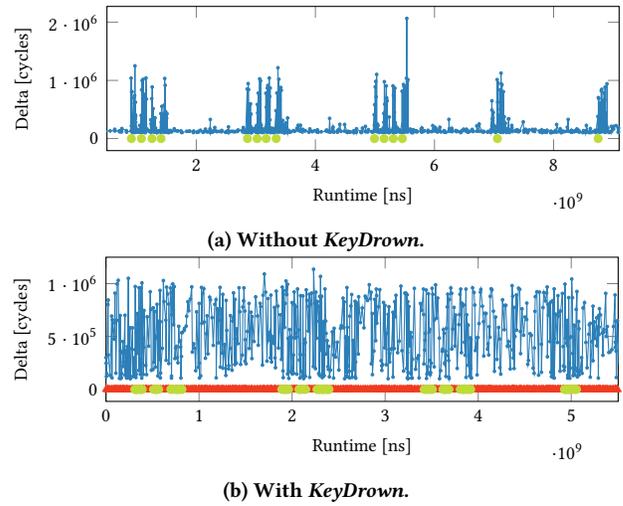
\begin{figure}[t]
\begin{subfigure}[t]{0.99\hsize}
 \centering
 \begin{tikzpicture}
\begin{axis}[
mlineplot,
style={font=\footnotesize},
xlabel={Runtime [ns]},
ylabel={Delta [cycles]},
width=1.0\hsize,
xmin=500000000,
xmax=9100000000,
scaled y ticks=false,
height=3.5cm
]

\addplot+[blue,thin,mark options={draw=blue,fill=blue},mark=*,mark size=0.5] table[x=Time,y=Value,col sep=comma] {data/layer1_rdtsc_without_keydrown_oneplus3t.csv};
\addplot+[only marks, mark options={draw=red,fill=red},mark=triangle*,mark size=1.5] table[x=Time,y=Fake,col sep=comma] {data/layer1_rdtsc_without_keydrown_oneplus3t.csv};
\addplot+[only marks,mark options={draw=green,fill=green},mark=*,mark size=1.5] table[x=Time,y=Real,col sep=comma] {data/layer1_rdtsc_without_keydrown_oneplus3t.csv};

\end{axis}
\end{tikzpicture}
 \caption{Without \KeyDrown.}
 \label{fig:layer1_rdtsc_wo_kd_oneplus3t}
\end{subfigure}
\begin{subfigure}[t]{0.99\hsize}
 \centering
 \begin{tikzpicture}
\begin{axis}[
mlineplot,
style={font=\footnotesize},
xlabel={Runtime [ns]},
ylabel={Delta [cycles]},
width=1.0\hsize,
xmin=0,
xmax=5504580000,
scaled y ticks=false,
height=3.5cm
]
\addplot+[blue,thin,mark options={draw=blue,fill=blue},mark=*,mark size=0.5] table[x=Time,y=Value,col sep=comma] {data/layer1_rdtsc_with_keydrown_oneplus3t.csv};
\addplot+[only marks, mark options={draw=red,fill=red},mark=triangle*,mark size=1.5] table[x=Time,y=Fake,col sep=comma] {data/layer1_rdtsc_with_keydrown_oneplus3t.csv};
\addplot+[only marks,mark options={draw=green,fill=green},mark=*,mark size=1.5] table[x=Time,y=Real,col sep=comma] {data/layer1_rdtsc_with_keydrown_oneplus3t.csv};


\end{axis}
\end{tikzpicture}
 \caption{With \KeyDrown.}
 \label{fig:layer1_rdtsc_w_kd_oneplus3t}
\end{subfigure}
\caption{Timing-based attack on the \OnePlus. (a) The attack allows to clearly detect every single
  touchscreen interrupt (\RealMarker). (b) With \KeyDrown, the attacker measures fake interrupts (\FakeMarker) as well as real interrupts (\RealMarker) and cannot distinguish
between them.}
\vspace{1.8cm}
\label{fig:rdtsc_attack_oneplus3t}
\end{figure}

Figure~\ref{fig:procfs_attack_oneplus3t} and Figure~\ref{fig:rdtsc_attack_oneplus3t} show a \texttt{procfs}-based interrupt attack as well as a timing-based attack, both on the \OnePlus. 
Without \KeyDrown, we achieve a precision of \SIx{\ValPrecisionProcOP} for the \texttt{procfs}-based attack and a precision of \SIx{\ValPrecisionRDTSCOP} for the timing-based attack, resulting in an \FScore of \SIx{\ValFscoreProcOP} and \SIx{\ValFscoreRDTSCOP} respectively. 
Enabling \KeyDrown reduces the precision to only \SIx{\ValPrecisionProcKD} and \SIx{\ValPrecisionRDTSCKD} respectively.
Thus, the resulting \FScore is \SIx{\ValFscoreProcKDOP} for the \texttt{procfs}-based attack, and \SIx{\ValFscoreRDTSCKDOP} for the timing-based attack, which is a \SI{0.0}{\percent} advantage over an always-one oracle.

\begin{figure}[!t]
    \centering
    \begin{subfigure}[t]{0.99\hsize}
      \centering
      \begin{tikzpicture}
\begin{axis}[
mlineplot,
style={font=\footnotesize},
xlabel={Runtime [ns]},
ylabel={Active cache sets},
width=1.0\hsize,
xmin=112800000,
xmax=2019611442,
ymax=6.5,
ymin=-1,
height=3.5cm
]
\addplot+[blue, no markers] table[x=Time,y=Value,col sep=comma] {data/layer1_prime_probe_without_keydrown_oneplus3t.csv};
\addplot+[only marks,mark options={draw=green,fill=green}, mark=*] table[x=Time,y=Real,col sep=comma] {data/layer1_prime_probe_without_keydrown_oneplus3t.csv};
\addplot+[only marks, mark options={draw=red,fill=red},mark=triangle*] table[x=Time,y=Fake,col sep=comma] {data/layer1_prime_probe_without_keydrown_oneplus3t.csv};

\end{axis}
\end{tikzpicture}
      \caption{Without \KeyDrown.}
      \label{fig:layer1_pp_kd_disabled_oneplus3t}
    \end{subfigure}%

    \begin{subfigure}[t]{0.99\hsize}
      \centering
      \begin{tikzpicture}
\begin{axis}[
mlineplot,
style={font=\footnotesize},
xlabel={Runtime [ns]},
ylabel={Active cache sets},
width=1.0\hsize,
xmin=0,
xmax=1343967790,
ymax=6.5,
ymin=-1,
height=3.5cm
]
\addplot+[blue, no markers] table[x=Time,y=Value,col sep=comma] {data/layer1_prime_probe_with_keydrown_oneplus3t.csv};
\addplot+[only marks, mark options={draw=red,fill=red},mark=triangle*] table[x=Time,y=Fake,col sep=comma] {data/layer1_prime_probe_with_keydrown_oneplus3t.csv};
\addplot+[only marks,mark options={draw=green,fill=green}, mark=*] table[x=Time,y=Real,col sep=comma] {data/layer1_prime_probe_with_keydrown_oneplus3t.csv};

\end{axis}
\end{tikzpicture}
      \caption{With \KeyDrown.}
      \label{fig:layer1_pp_kd_oneplus3t}
    \end{subfigure}%

    \caption{\MultiPrimeProbe attack on the \SIx{\ValPPCacheSets} cache sets from \texttt{0x3fc0355c28} to \texttt{0x3fc0355d68} of \texttt{msm\_gpio\_irq\_handler} of the \OnePlus. 
    (a) Noise negatively affects the detection of single keystrokes (\RealMarker). 
    (b) With \KeyDrown, the attacker measures even more cache misses on injected keystrokes (\FakeMarker) as well as on real events (\RealMarker) and cannot distinguish between them.}
    \label{fig:layer1_pp_kernel_oneplus3t}
\end{figure}
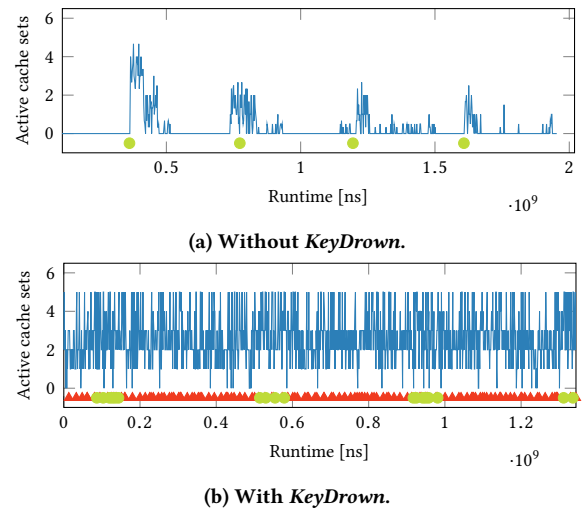

Figure~\ref{fig:layer1_pp_kernel_oneplus3t} shows the results of inferring keystroke timings by detecting the touchscreen interrupt handler's cache activity using \MultiPrimeProbe on the \OnePlus.
We monitored \SIx{\ValPPCacheSetsOP} cache sets in parallel for a higher noise robustness.
Without \KeyDrown, the precision is already at a quite low value of \SIx{\ValPrecisionPPOP} with a recall of only \SIx{\ValRecallPPOP}, as an access to one of the cache sets by any other application cannot be distinguished from a cache set access by the touchscreen interrupt handler.
Thus, this attack has a high number of false positives. 
If we enable \KeyDrown, the precision drops to \SIx{\ValPrecisionPPKDOP}, as the attacker additionally measures the noise generated by the injected keystrokes. 
Thus, the \FScore is \SIx{\ValFscorePPKDOP}, which is a $\leq$\SI{-52.7}{\percent} advantage over an always-one oracle.

\end{document}